\documentclass[10pt,a4paper]{article}
\usepackage{amsmath,amssymb,amsfonts}
\usepackage{graphicx}
\usepackage{hyperref}
\usepackage[margin=0.75in]{geometry}
\usepackage{float}
\usepackage{bm}
\usepackage{booktabs}
\usepackage{tabularx}

\usepackage{setspace}
\usepackage{titlesec}
\titlespacing*{\section}{0pt}{8pt}{4pt}
\titlespacing*{\subsection}{0pt}{6pt}{3pt}

\AtBeginDocument{
  \setlength{\abovedisplayskip}{4pt}
  \setlength{\belowdisplayskip}{4pt}
  \setlength{\abovedisplayshortskip}{2pt}
  \setlength{\belowdisplayshortskip}{2pt}
}

\let\oldbibliography\thebibliography
\renewcommand{\thebibliography}[1]{%
  \oldbibliography{#1}%
  \setlength{\itemsep}{0pt}%
  \setlength{\parskip}{0pt}%
}

\title{An Open-Source Pseudo-Spectral Solver for Idealized
Korteweg--de Vries Soliton Simulations}

\author{
Dasapta Erwin Irawan$^{1,*}$,
Sandy Hardian Susanto Herho$^{2,3,4,5}$,
Astyka Pamumpuni$^{1}$,\\
Rendy Dwi Kartiko$^{1}$,
Faruq Khadami$^{6}$,
Iwan Pramesti Anwar$^{6}$,\\
Karina Aprilia Sujatmiko$^{6}$,
Alfita Puspa Handayani$^{5,7}$,\\
Faiz Rohman Fajary$^{8}$,
and Rusmawan Suwarman$^{8}$
}

\date{}

\begin{document}
\maketitle

\begin{center}
\small
$^{1}$Applied Geology Research Group, Bandung Institute of Technology,
Bandung, West Java 40132, Indonesia\\
$^{2}$Department of Earth and Planetary Sciences, University of California,
Riverside, CA 92521, USA\\
$^{3}$Ronin Institute for Independent Scholarship 2.0,
Sacramento, CA 95816, USA\\
$^{4}$School of Systems Science and Industrial Engineering,
State University of New York, Binghamton, NY 13902, USA\\
$^{5}$Center for Agrarian Studies, Bandung Institute of Technology,
Bandung, West Java 40132, Indonesia\\
$^{6}$Applied and Environmental Oceanography Research Group,
Bandung Institute of Technology, Bandung, West Java 40132, Indonesia\\
$^{7}$Spatial Systems and Cadaster Research Group,
Bandung Institute of Technology, Bandung, West Java 40132, Indonesia\\
$^{8}$Atmospheric Science Research Group, Bandung Institute of Technology,
Bandung, West Java 40132, Indonesia\\
$^{*}$Correspondence: dasaptaerwin@itb.ac.id
\end{center}

\begin{abstract}
\noindent
The Korteweg--de Vries (KdV) equation is a foundational model in
geophysical fluid dynamics (GFD), governing the propagation of long
internal and surface gravity waves in stratified and shallow ocean
environments where the interplay between nonlinear steepening and
frequency-dependent dispersion gives rise to solitons. Although the
analytical tractability of the KdV equation through inverse scattering
is well established, systematic numerical exploration of multi-soliton
interactions remains valuable for benchmarking solvers, probing
conservation properties under varied oceanic initial conditions, and
building intuition for more complex ocean wave phenomena. This article
presents \texttt{sangkuriang}, an open-source Python library that solves
the KdV equation using Fourier pseudo-spectral spatial discretization and
adaptive eighth-order Runge--Kutta time integration. The implementation
leverages just-in-time (JIT) compilation to achieve research-grade
computational efficiency on standard hardware. The solver is validated
through four progressively complex idealized scenarios motivated by
oceanic wave dynamics: isolated soliton propagation, symmetric
interactions, overtaking collisions, and three-body interactions.
High-fidelity conservation of mass, momentum, and energy is demonstrated,
with relative errors remaining below $\mathcal{O}(10^{-4})$ across all
test cases. Measured soliton velocities align with theoretical predictions
within 5\%, confirming the capture of the amplitude-dependent dispersion
characteristic of oceanic solitary waves. Complementary diagnostics,
including spectral entropy and recurrence quantification analysis (RQA),
verify that the numerical solutions preserve the regular phase-space
structure characteristic of integrable Hamiltonian systems. These results
establish \texttt{sangkuriang} as a robust, lightweight platform for
reproducible numerical investigation of idealized nonlinear dispersive
wave dynamics relevant to coastal and ocean engineering applications.
\end{abstract}


\section{Introduction}

The Korteweg--de Vries (KdV) equation occupies a central position in
geophysical fluid dynamics (GFD) as the prototypical completely integrable
partial differential equation (PDE) admitting soliton
solutions~\cite{ref-1}. Originally derived to describe long gravity waves
in shallow channels~\cite{ref-2}, the equation demonstrates how competing
nonlinear steepening and linear dispersion can balance to produce
localized, shape-preserving pulses---solitons---that survive mutual
collisions with only phase shifts~\cite{ref-3}. The discovery that this
behavior is rooted in complete integrability, with an infinite hierarchy
of conservation laws accessible through inverse
scattering~\cite{ref-4,ref-5}, elevated the KdV equation to a cornerstone
of mathematical physics, with applications spanning plasma dynamics,
internal ocean waves, and condensed matter systems~\cite{ref-6}.

In the oceanic context, the KdV framework and its extensions have proven
indispensable for describing nonlinear internal solitary waves (ISWs),
which are among the most energetic high-frequency phenomena in the coastal
and deep ocean~\cite{ref-46,ref-47}. Satellite imagery and in situ
observations have documented large-amplitude ISWs in numerous marginal
seas, including the South China Sea, the Andaman Sea~\cite{ref-49}, and
the Sulu Sea, where these waves modulate acoustic propagation, drive
turbulent mixing, and exert forces on offshore structures~\cite{ref-48}.
While realistic ocean modeling requires variable-coefficient and
higher-order extensions of the KdV equation~\cite{ref-48,ref-50}, the
constant-coefficient idealization retains its importance as the canonical
benchmark for understanding the fundamental balance between nonlinearity
and dispersion that underpins oceanic solitary wave
dynamics~\cite{ref-6,ref-46}.

Although the analytical structure of soliton theory is well established,
numerical simulation remains indispensable for exploring parameter
regimes, validating approximate initial conditions, and characterizing
transient interaction dynamics that resist closed-form
description~\cite{ref-7}. In particular, the quantitative assessment of
conservation law preservation, amplitude-dependent dispersion, and
phase-space regularity across multi-soliton configurations requires
flexible computational tools capable of high-fidelity integration over
extended time intervals~\cite{ref-8}. Such capabilities are especially
relevant to idealized studies of oceanic internal wave interactions, where
multi-soliton collisions and overtaking events are routinely observed in
field campaigns~\cite{ref-51}.

Python has become a widely adopted language for scientific computing in
the geosciences and oceanography, owing to its extensive numerical
ecosystem and low overhead for prototyping~\cite{ref-9,ref-10}. Libraries
such as NumPy and SciPy~\cite{ref-11,ref-12} provide efficient
implementations of core algorithms---fast Fourier transforms (FFT),
adaptive ODE solvers, and dense linear algebra---while just-in-time (JIT)
compilation via Numba~\cite{ref-13} bridges the gap between interpreted
flexibility and compiled performance~\cite{ref-14}. These capabilities
have made Python increasingly viable for production-level numerical
experiments in computational geophysics~\cite{ref-15,ref-16}.

Pseudo-spectral methods are particularly well suited to the KdV equation.
By representing spatial derivatives as multiplications in wavenumber
space, Fourier-based discretizations achieve exponential convergence for
smooth periodic solutions while naturally respecting the spectral
structure of dispersive operators~\cite{ref-17,ref-18}. This approach has
a long and successful history in the numerical treatment of nonlinear
dispersive PDEs arising in ocean wave theory~\cite{ref-19,ref-20}.

In this work, we present \texttt{sangkuriang}, an open-source Python
library for solving the KdV equation using Fourier pseudo-spectral spatial
discretization coupled with adaptive eighth-order Runge--Kutta time
integration via the Dormand--Prince 8(5,3) (DOP853)
method~\cite{ref-21,ref-22}. Performance-critical routines are accelerated
through Numba JIT compilation with multi-core parallelization, enabling
research-grade simulations on commodity hardware. The library provides
both a command-line interface for rapid deployment with predefined test
cases and a Python API for programmatic access. Simulation data are output
in Network Common Data Format (NetCDF) following Climate and Forecast (CF)
conventions~\cite{ref-23}, ensuring interoperability with standard
oceanographic and geoscientific analysis workflows. The solver additionally
incorporates diagnostic routines for monitoring conservation laws, tracking
soliton trajectories, and generating animated visualizations. The software
is freely available from the Python Package Index (PyPI) under a permissive
open-source license.

We revisit the classical benchmarks of KdV soliton dynamics---single-soliton
propagation, symmetric two-soliton configurations, overtaking collisions,
and three-body interactions---framed as idealized analogs of oceanic
solitary wave phenomena, with an emphasis on quantitative diagnostics that
go beyond visual inspection. Specifically, we employ information-theoretic
measures (spectral entropy, LMC statistical complexity, Fisher information)
and recurrence quantification analysis to characterize the phase-space
structure of the computed solutions. These complementary diagnostics provide
model-independent evidence for the preservation of integrability under
numerical integration and offer tools that may prove useful in studying
perturbed or extended KdV-type systems relevant to realistic ocean
stratification and bathymetric variability, where analytical guarantees are
unavailable.

\section{Materials and Methods}

\subsection{Model Description}

The KdV equation constitutes one of the most fundamental nonlinear PDEs in
GFD, governing the propagation of weakly nonlinear, weakly dispersive long
waves in shallow ocean basins, coastal shelves, and stratified water
columns~\cite{ref-2,ref-3}. The derivation presented here proceeds from
first principles beginning with the three-dimensional Euler equations for
an inviscid ocean layer, following the established asymptotic analysis
framework employed in ocean wave theory~\cite{ref-7,ref-6}.

\begin{figure}[H]
\centering
\includegraphics[width=0.8\textwidth]{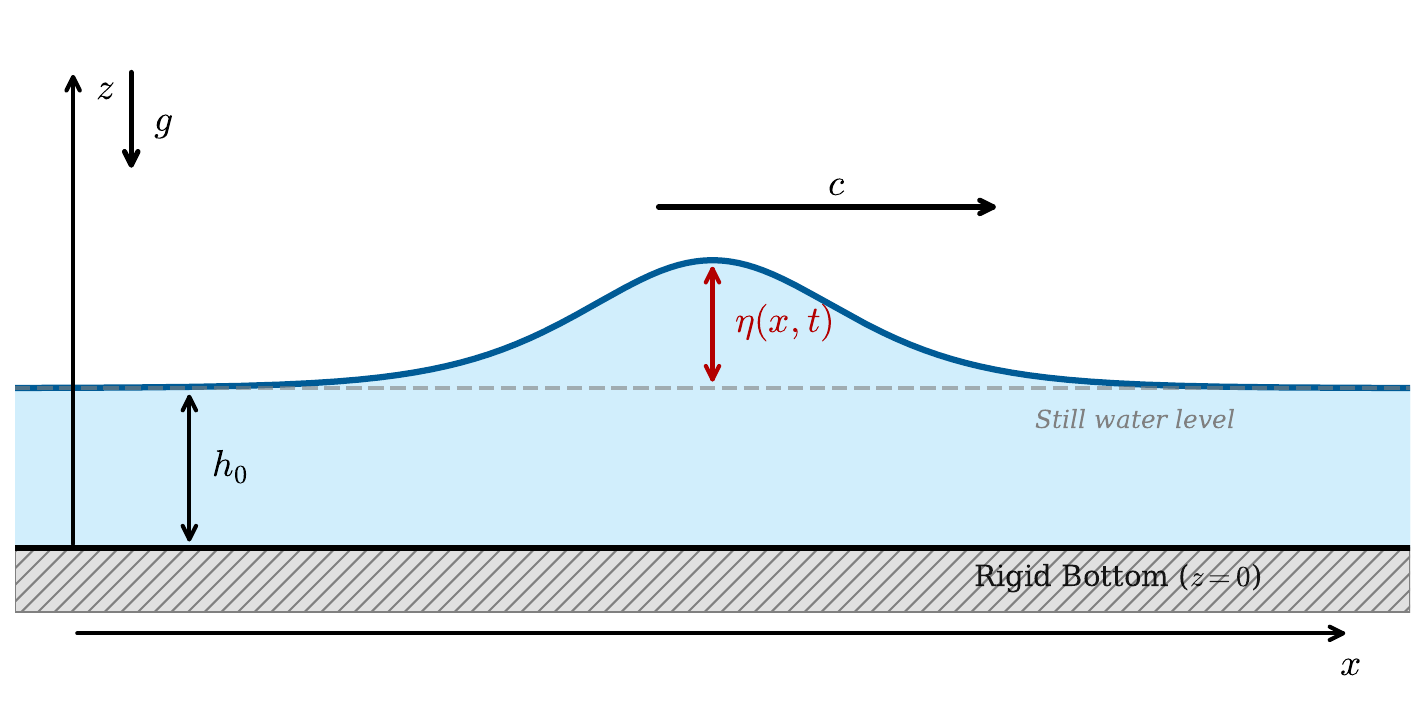}
\caption{Schematic of the shallow ocean wave domain. The fluid occupies a
region bounded below by a rigid horizontal seabed at $z = 0$ and above by
a free surface at $z = h_0 + \eta(x,t)$, where $h_0$ denotes the
undisturbed water depth and $\eta(x,t)$ represents the free surface
displacement. Gravitational acceleration $g$ acts in the negative
$z$-direction, and wave propagation occurs along the $x$-axis with phase
speed $c$.}
\label{fig:domain_schematic}
\end{figure}

Consider an inviscid, incompressible fluid occupying a domain bounded below
by a rigid horizontal seabed at $z = 0$ and above by a free surface at
$z = h_0 + \eta(x,y,t)$, where $h_0$ denotes the undisturbed water depth
and $\eta$ represents the free surface displacement
(Figure~\ref{fig:domain_schematic}). The fluid is subject to gravitational
acceleration $g$ acting in the negative $z$-direction. This configuration
represents an idealized coastal or continental shelf environment in which
long surface gravity waves propagate over a flat bottom---a standard
starting point for shallow water wave theory in physical
oceanography~\cite{ref-24,ref-46}.

For an inviscid fluid, the three-dimensional incompressible Euler equations
governing the velocity field $\mathbf{u} = (u, v, w)$ and pressure $p$
read
\begin{equation}
\nabla \cdot \mathbf{u} = \frac{\partial u}{\partial x}
+ \frac{\partial v}{\partial y} + \frac{\partial w}{\partial z} = 0,
\label{eq:continuity_3d}
\end{equation}
\begin{equation}
\frac{\partial u}{\partial t} + u\frac{\partial u}{\partial x}
+ v\frac{\partial u}{\partial y} + w\frac{\partial u}{\partial z}
= -\frac{1}{\rho}\frac{\partial p}{\partial x},
\label{eq:momentum_x}
\end{equation}
\begin{equation}
\frac{\partial v}{\partial t} + u\frac{\partial v}{\partial x}
+ v\frac{\partial v}{\partial y} + w\frac{\partial v}{\partial z}
= -\frac{1}{\rho}\frac{\partial p}{\partial y},
\label{eq:momentum_y}
\end{equation}
\begin{equation}
\frac{\partial w}{\partial t} + u\frac{\partial w}{\partial x}
+ v\frac{\partial w}{\partial y} + w\frac{\partial w}{\partial z}
= -\frac{1}{\rho}\frac{\partial p}{\partial z} - g,
\label{eq:momentum_z}
\end{equation}
where $\rho$ is the constant fluid density. These equations are
supplemented by boundary conditions at the seabed and free
surface~\cite{ref-24,ref-25}.

At the rigid seabed $z = 0$, the impermeability condition requires
\begin{equation}
w = 0 \quad \text{at} \quad z = 0.
\label{eq:bottom_bc}
\end{equation}
At the free surface $z = h_0 + \eta(x,y,t)$, the kinematic boundary
condition states that fluid particles on the surface remain on the surface:
\begin{equation}
w = \frac{\partial \eta}{\partial t} + u\frac{\partial \eta}{\partial x}
+ v\frac{\partial \eta}{\partial y}
\quad \text{at} \quad z = h_0 + \eta,
\label{eq:kinematic_bc}
\end{equation}
and the dynamic boundary condition requires continuity of normal stress,
which for an inviscid fluid with negligible surface tension reduces to
\begin{equation}
p = p_{\text{atm}} \quad \text{at} \quad z = h_0 + \eta,
\label{eq:dynamic_bc}
\end{equation}
where $p_{\text{atm}}$ is the constant atmospheric pressure.

We now restrict attention to unidirectional wave propagation along the
$x$-axis, setting $v = 0$ and $\partial/\partial y = 0$. This reduction
is appropriate for long-crested ocean waves propagating shoreward over a
continental shelf or along a waveguide~\cite{ref-47}. The governing
equations~\eqref{eq:continuity_3d}--\eqref{eq:momentum_z} reduce to the
two-dimensional system
\begin{equation}
\frac{\partial u}{\partial x} + \frac{\partial w}{\partial z} = 0,
\label{eq:continuity_2d}
\end{equation}
\begin{equation}
\frac{\partial u}{\partial t} + u\frac{\partial u}{\partial x}
+ w\frac{\partial u}{\partial z}
= -\frac{1}{\rho}\frac{\partial p}{\partial x},
\label{eq:momentum_2d_x}
\end{equation}
\begin{equation}
\frac{\partial w}{\partial t} + u\frac{\partial w}{\partial x}
+ w\frac{\partial w}{\partial z}
= -\frac{1}{\rho}\frac{\partial p}{\partial z} - g.
\label{eq:momentum_2d_z}
\end{equation}

To derive the shallow water equations, we introduce characteristic scales:
$\lambda$ for horizontal length (the dominant wavelength), $h_0$ for
vertical length (the water depth), $a$ for wave amplitude, $\lambda/c_0$
for time (where $c_0 = \sqrt{gh_0}$ is the linear long-wave speed), $c_0$
for horizontal velocity, and $\rho g h_0$ for pressure. The dimensionless
variables are defined as
\begin{equation}
\tilde{x} = \frac{x}{\lambda}, \quad \tilde{z} = \frac{z}{h_0}, \quad
\tilde{t} = \frac{c_0 t}{\lambda}, \quad \tilde{\eta} = \frac{\eta}{a},
\quad \tilde{u} = \frac{u}{c_0}, \quad
\tilde{w} = \frac{w\lambda}{c_0 h_0}, \quad
\tilde{p} = \frac{p - p_{\text{atm}}}{\rho g h_0}.
\label{eq:nondim_vars}
\end{equation}
Two fundamental small parameters emerge naturally: the amplitude parameter
$\alpha = a/h_0 \ll 1$ measuring wave nonlinearity, and the dispersion
parameter $\beta = (h_0/\lambda)^2 \ll 1$ quantifying the shallowness of
the water~\cite{ref-26}. In typical continental shelf settings where long
gravity waves or internal solitary waves are observed, both $\alpha$ and
$\beta$ are indeed small, placing the dynamics squarely within the weakly
nonlinear, weakly dispersive regime that the KdV equation
governs~\cite{ref-46,ref-48}.

Substituting the scalings~\eqref{eq:nondim_vars} into
equations~\eqref{eq:continuity_2d}--\eqref{eq:momentum_2d_z} and dropping
tildes for clarity, we obtain the dimensionless system
\begin{equation}
\frac{\partial u}{\partial x} + \frac{\partial w}{\partial z} = 0,
\label{eq:nondim_continuity}
\end{equation}
\begin{equation}
\frac{\partial u}{\partial t} + \alpha\left(u\frac{\partial u}{\partial x}
+ w\frac{\partial u}{\partial z}\right) = -\frac{\partial p}{\partial x},
\label{eq:nondim_mom_x}
\end{equation}
\begin{equation}
\beta\left[\frac{\partial w}{\partial t}
+ \alpha\left(u\frac{\partial w}{\partial x}
+ w\frac{\partial w}{\partial z}\right)\right]
= -\frac{\partial p}{\partial z} - 1.
\label{eq:nondim_mom_z}
\end{equation}
The factor $\beta$ multiplying the left-hand side of
equation~\eqref{eq:nondim_mom_z} indicates that vertical accelerations are
negligible in the shallow water limit $\beta \to 0$, consistent with the
hydrostatic approximation widely employed in large-scale ocean circulation
models~\cite{ref-27}.

In the leading-order shallow water approximation ($\beta = 0$),
equation~\eqref{eq:nondim_mom_z} reduces to the hydrostatic balance
\begin{equation}
\frac{\partial p}{\partial z} = -1,
\label{eq:hydrostatic}
\end{equation}
which, upon integration from $z$ to the free surface $z = 1 + \alpha\eta$
using the dynamic boundary condition $p = 0$ at $z = 1 + \alpha\eta$,
yields
\begin{equation}
p = 1 + \alpha\eta - z.
\label{eq:pressure_hydrostatic}
\end{equation}
Thus, the horizontal pressure gradient becomes
\begin{equation}
\frac{\partial p}{\partial x} = \alpha\frac{\partial \eta}{\partial x},
\label{eq:pressure_gradient}
\end{equation}
independent of $z$, implying that the horizontal velocity $u$ is also
independent of depth at leading order.

Integrating the continuity equation~\eqref{eq:nondim_continuity} from
$z = 0$ to $z = 1 + \alpha\eta$ and applying the boundary
conditions~\eqref{eq:bottom_bc} and~\eqref{eq:kinematic_bc}, we obtain
\begin{equation}
\frac{\partial \eta}{\partial t}
+ \frac{\partial}{\partial x}\left[(1 + \alpha\eta)u\right] = 0.
\label{eq:mass_conservation}
\end{equation}
Combining with the horizontal momentum
equation~\eqref{eq:nondim_mom_x} and pressure
gradient~\eqref{eq:pressure_gradient}, we arrive at the nonlinear shallow
water equations in dimensionless form~\cite{ref-24}
\begin{equation}
\frac{\partial \eta}{\partial t}
+ \frac{\partial}{\partial x}\left[(1 + \alpha\eta)u\right] = 0,
\label{eq:shallow_mass}
\end{equation}
\begin{equation}
\frac{\partial u}{\partial t} + \alpha u\frac{\partial u}{\partial x}
+ \frac{\partial \eta}{\partial x} = 0.
\label{eq:shallow_momentum}
\end{equation}
These equations are exact within the hydrostatic shallow water
approximation but remain coupled and nonlinear. They form the basis of
many operational coastal and tsunami models, but lack the dispersive
corrections necessary to sustain solitary wave solutions~\cite{ref-50}.

To incorporate weak dispersion and derive the KdV equation, we retain
terms of order $\beta$ in the vertical momentum
equation~\eqref{eq:nondim_mom_z}. The regime
$\alpha \sim \beta \sim \mathcal{O}(\delta) \ll 1$ with $\delta$
representing a single small parameter defines the Boussinesq scaling,
wherein nonlinear and dispersive effects balance to produce solitary wave
solutions~\cite{ref-26}. This balance is the essential physical mechanism
underlying the formation of internal solitary waves observed in
shelf-slope regions worldwide~\cite{ref-46,ref-49}. Under this scaling,
we seek asymptotic solutions in the form of perturbation expansions
\begin{equation}
\eta(\xi,\tau) = \delta\eta_1(\xi,\tau)
+ \delta^2\eta_2(\xi,\tau) + \mathcal{O}(\delta^3),
\label{eq:eta_expansion}
\end{equation}
\begin{equation}
u(\xi,\tau) = \delta u_1(\xi,\tau)
+ \delta^2 u_2(\xi,\tau) + \mathcal{O}(\delta^3),
\label{eq:u_expansion}
\end{equation}
where $\delta = \sqrt{\alpha} = \sqrt{\beta}$ serves as the expansion
parameter. Substituting
expansions~\eqref{eq:eta_expansion} and~\eqref{eq:u_expansion} into the
shallow water equations and collecting terms of equal powers in $\delta$,
we obtain at leading order $\mathcal{O}(\delta)$ the linearized wave
equations
\begin{equation}
\frac{\partial \eta_1}{\partial \tau}
+ \frac{\partial u_1}{\partial \xi} = 0,
\label{eq:linear_mass}
\end{equation}
\begin{equation}
\frac{\partial u_1}{\partial \tau}
+ \frac{\partial \eta_1}{\partial \xi} = 0.
\label{eq:linear_momentum}
\end{equation}
These admit right-going wave solutions of the form
$\eta_1 = f(\xi - \tau)$ and $u_1 = f(\xi - \tau)$, where $f$ is an
arbitrary waveform propagating with dimensionless speed unity (dimensional
speed $c_0 = \sqrt{gh_0}$).

At the next order $\mathcal{O}(\delta^2)$, nonlinear and dispersive
corrections emerge~\cite{ref-1}:
\begin{equation}
\frac{\partial \eta_2}{\partial \tau}
+ \frac{\partial u_2}{\partial \xi}
= -\frac{\partial}{\partial \xi}(\eta_1 u_1),
\label{eq:second_mass}
\end{equation}
\begin{equation}
\frac{\partial u_2}{\partial \tau}
+ \frac{\partial \eta_2}{\partial \xi}
= -u_1\frac{\partial u_1}{\partial \xi}
+ \frac{1}{3}\frac{\partial^3 u_1}{\partial \xi^2 \partial \tau}.
\label{eq:second_momentum}
\end{equation}
The nonlinear term $-\partial_\xi(\eta_1 u_1)$ in
equation~\eqref{eq:second_mass} arises from the advection of mass by the
wave motion, while the dispersive term
$\frac{1}{3}\partial_\xi^2\partial_\tau u_1$ in
equation~\eqref{eq:second_momentum} originates from the vertical
acceleration induced by the sloping bottom pressure
distribution~\cite{ref-27}.

To eliminate secular growth and derive a consistent evolution equation,
we introduce the stretched slow variables $X = \delta(\xi - \tau)$ and
$T = \delta^2\tau$, representing a coordinate system moving with the
linear wave speed and a slow time scale appropriate for observing
cumulative nonlinear and dispersive effects~\cite{ref-6}. We seek
solutions in the form $\eta_1 = U(X,T)$ and apply the multiple scales
method. The chain rule yields
\begin{equation}
\frac{\partial}{\partial \xi} = \delta\frac{\partial}{\partial X}, \quad
\frac{\partial}{\partial \tau} = -\delta\frac{\partial}{\partial X}
+ \delta^2\frac{\partial}{\partial T}.
\label{eq:chain_rule}
\end{equation}

Combining the solvability conditions at order $\mathcal{O}(\delta^3)$ to
suppress resonant secular terms, and transforming back to dimensional
variables through the scalings $x = \lambda X$,
$t = \lambda T\sqrt{h_0/g}$, and $\eta = aU$, we obtain the canonical KdV
equation in dimensional form~\cite{ref-7}
\begin{equation}
\frac{\partial \eta}{\partial t}
+ c_0\frac{\partial \eta}{\partial x}
+ \frac{3c_0}{2h_0}\eta\frac{\partial \eta}{\partial x}
+ \frac{c_0h_0^2}{6}\frac{\partial^3 \eta}{\partial x^3} = 0,
\label{eq:kdv_dimensional}
\end{equation}
where $c_0 = \sqrt{gh_0}$ denotes the linear shallow water wave speed.
This formulation makes explicit the physical origins of each term: the
second term represents linear wave propagation at the long-wave phase
speed, the third term captures amplitude-dependent steepening
(nonlinearity) responsible for wave breaking in the absence of dispersion,
and the fourth term describes frequency-dependent dispersion that spreads
energy across wavenumbers. The competition between these last two effects
is the fundamental mechanism sustaining solitary waves in shallow ocean
environments~\cite{ref-46,ref-48}.

For computational implementation and to emphasize the balance between
nonlinearity and dispersion, we transform
equation~\eqref{eq:kdv_dimensional} to a traveling coordinate frame
$\chi = x - c_0t$ and define $u(\chi,t) = \eta(x,t)$. Furthermore, we
introduce dimensionless parameters $\varepsilon = 3c_0/(2h_0)$
[m$^{-1}$] characterizing the nonlinearity strength and
$\mu = c_0h_0^2/6$ [m$^3$/s] quantifying dispersion. This yields the
standard form employed in the numerical solver~\cite{ref-17,ref-20}
\begin{equation}
\frac{\partial u}{\partial t}
+ \varepsilon u\frac{\partial u}{\partial x}
+ \mu\frac{\partial^3 u}{\partial x^3} = 0,
\label{eq:kdv_standard}
\end{equation}
where we have relabeled $\chi \to x$ for notational simplicity. In
equation~\eqref{eq:kdv_standard}, $u(x,t)$ represents the wave amplitude
field [m], with spatial coordinate $x$ [m] and temporal coordinate $t$
[s]. The remarkable property of this equation lies in the exact balance
between the quadratic nonlinear term $\varepsilon u\partial_x u$, which
steepens wave profiles and would induce shock formation in isolation, and
the third-order dispersive term $\mu\partial_x^3 u$, which spreads energy
and favors shorter wavelengths. When these mechanisms achieve equilibrium,
the equation admits solitary wave solutions---solitons---that propagate
without change of form~\cite{ref-3}. This idealized balance captures the
essential physics of long nonlinear waves observed on continental shelves,
in straits, and in the thermocline of the open ocean~\cite{ref-47,ref-51}.

The KdV equation~\eqref{eq:kdv_standard} possesses profound mathematical
structure as a completely integrable Hamiltonian system. It admits an
infinite hierarchy of conservation laws, the first three of which
correspond to physically meaningful quantities~\cite{ref-28,ref-4}. The
mass functional
\begin{equation}
M[u] = \int_{-\infty}^{\infty} u(x,t)\,dx
\label{eq:mass_invariant}
\end{equation}
represents the total wave amplitude integrated over space, analogous to
volume conservation in the ocean surface layer. The momentum functional
\begin{equation}
P[u] = \int_{-\infty}^{\infty} u^2(x,t)\,dx
\label{eq:momentum_invariant}
\end{equation}
corresponds to the square-integrated field, related to wave action in the
physical system. The energy functional
\begin{equation}
E[u] = \int_{-\infty}^{\infty} \left[\frac{\varepsilon}{2}u^3
- \frac{3\mu}{2}\left(\frac{\partial u}{\partial x}\right)^2\right]dx
\label{eq:energy_invariant}
\end{equation}
represents the Hamiltonian of the system, with the cubic term
$\frac{\varepsilon}{2}u^3$ encoding nonlinear potential energy and the
quadratic gradient term $-\frac{3\mu}{2}(\partial_x u)^2$ representing
dispersive kinetic energy. These
invariants~\eqref{eq:mass_invariant}--\eqref{eq:energy_invariant} satisfy
$dM/dt = 0$, $dP/dt = 0$, and $dE/dt = 0$ along solutions of
equation~\eqref{eq:kdv_standard}, providing essential diagnostics for
numerical accuracy~\cite{ref-6}. Their preservation under numerical
integration is a necessary condition for any solver intended to reproduce
the long-time dynamics of oceanic solitary waves faithfully.

The KdV equation~\eqref{eq:kdv_standard} admits exact single-soliton
solutions in the form~\cite{ref-1}
\begin{equation}
u(x,t) = \frac{A}{\cosh^2\left[\sqrt{\frac{\varepsilon A}{12\mu}}
(x - x_0 - vt)\right]},
\label{eq:soliton_solution}
\end{equation}
where $A > 0$ denotes the soliton amplitude [m], $x_0$ represents the
initial position [m], and the propagation velocity [m/s] is given by
\begin{equation}
v = \frac{\varepsilon A}{3}.
\label{eq:soliton_velocity}
\end{equation}
Equation~\eqref{eq:soliton_velocity} reveals the fundamental nonlinear
dispersion relation: taller solitons travel faster, enabling overtaking
collisions. This amplitude--velocity relationship is a robust feature of
oceanic internal solitary waves, where larger-amplitude ISWs are
consistently observed to propagate faster than their smaller
counterparts~\cite{ref-46,ref-49}. The width parameter
$w = \sqrt{12\mu/(\varepsilon A)}$ [m] characterizes the spatial extent,
demonstrating that larger amplitudes produce narrower, faster solitons---a
relationship well documented in field observations of ISWs on continental
shelves~\cite{ref-48}. The remarkable stability of these solutions under
mutual collisions---emerging unchanged except for phase shifts---constitutes
the defining characteristic of soliton behavior and validates the KdV
equation as an integrable system~\cite{ref-3,ref-5}.

\subsection{Numerical Implementation}

The numerical solution of the KdV
equation~\eqref{eq:kdv_standard} requires spatial discretization of the
differential operators and temporal integration of the resulting
semi-discrete system. The \texttt{sangkuriang} solver employs the Fourier
pseudo-spectral method for spatial derivatives~\cite{ref-18,ref-19}, which
may provide exponential convergence rates for smooth periodic solutions,
combined with adaptive high-order explicit time integration. This approach
has been demonstrated to be effective for nonlinear dispersive wave
equations arising in ocean and atmospheric dynamics~\cite{ref-17,ref-8}.

Consider a periodic spatial domain $x \in [x_{\min}, x_{\max}]$ with
length $L = x_{\max} - x_{\min}$. We discretize this domain using $N$
uniformly spaced grid points $x_j = x_{\min} + j\Delta x$ for
$j = 0, 1, \ldots, N-1$, where the spatial resolution is $\Delta x = L/N$.
Let $u_j(t) = u(x_j, t)$ denote the semi-discrete solution at grid point
$x_j$ and time $t$. The periodicity condition
$u(x_{\min}, t) = u(x_{\max}, t)$ is implicitly enforced through the DFT
representation. While ocean domains are not strictly periodic, the periodic
boundary assumption is standard in idealized KdV studies and is justified
when soliton amplitudes decay sufficiently before reaching the domain
boundaries~\cite{ref-17,ref-20}. The DFT and its inverse are defined
as~\cite{ref-29}
\begin{equation}
\hat{u}_k = \frac{1}{N}\sum_{j=0}^{N-1} u_j
\exp\left(-\frac{2\pi i k j}{N}\right),
\quad k = 0, 1, \ldots, N-1,
\label{eq:dft}
\end{equation}
\begin{equation}
u_j = \sum_{k=0}^{N-1} \hat{u}_k
\exp\left(\frac{2\pi i k j}{N}\right),
\quad j = 0, 1, \ldots, N-1,
\label{eq:idft}
\end{equation}
where $\hat{u}_k$ represents the $k$-th Fourier coefficient and
$i = \sqrt{-1}$ denotes the imaginary unit. The computational
implementation utilizes the FFT algorithm~\cite{ref-30} available in
NumPy~\cite{ref-11}, reducing the operation count from $\mathcal{O}(N^2)$
to $\mathcal{O}(N\log N)$ for both forward and inverse transforms.

The wavenumber vector corresponding to the DFT basis functions is given by
\begin{equation}
k_n = \begin{cases}
\dfrac{2\pi n}{L}, & n = 0, 1, \ldots, \lfloor N/2 \rfloor, \\[8pt]
\dfrac{2\pi(n - N)}{L}, & n = \lceil N/2 \rceil, \ldots, N-1,
\end{cases}
\label{eq:wavenumber}
\end{equation}
where the convention in equation~\eqref{eq:wavenumber} accounts for the
standard FFT frequency ordering, with negative frequencies wrapped to the
upper half of the index range. This wavenumber array can be efficiently
generated using \texttt{numpy.fft.fftfreq}~\cite{ref-11}. The spectral
derivative operator acts on the Fourier coefficients through multiplication
by $ik_n$. For the first-order spatial derivative, we have
\begin{equation}
\frac{\partial u}{\partial x}(x_j) =
\text{IFFT}\left[ik_n \cdot \text{FFT}[u_j]\right]
= \text{IFFT}[ik_n \hat{u}_k],
\label{eq:first_derivative}
\end{equation}
where $\text{FFT}[\cdot]$ and $\text{IFFT}[\cdot]$ denote the
DFT~\eqref{eq:dft} and its inverse~\eqref{eq:idft}, respectively. The
third-order derivative required for the dispersion term follows
analogously:
\begin{equation}
\frac{\partial^3 u}{\partial x^3}(x_j)
= \text{IFFT}[(ik_n)^3 \hat{u}_k].
\label{eq:third_derivative}
\end{equation}
The real part of the inverse transform is extracted to eliminate numerical
round-off errors in the imaginary component, which should theoretically
vanish for real-valued solutions. This spectral differentiation approach is
exact in the continuous limit and introduces errors only through aliasing
in the nonlinear term, which can be controlled through sufficient spatial
resolution~\cite{ref-31}.

Applying the spatial discretization described by
equations~\eqref{eq:first_derivative} and~\eqref{eq:third_derivative} to
the KdV equation~\eqref{eq:kdv_standard}, we obtain the semi-discrete
ordinary differential equation (ODE) system
\begin{equation}
\frac{du_j}{dt} = F_j(\mathbf{u}(t)),
\quad j = 0, 1, \ldots, N-1,
\label{eq:semi_discrete}
\end{equation}
where $\mathbf{u}(t) = [u_0(t), u_1(t), \ldots, u_{N-1}(t)]^T
\in \mathbb{R}^N$ denotes the state vector and the right-hand side
operator is defined as
\begin{equation}
F_j(\mathbf{u}) = -\varepsilon u_j
\left[\frac{\partial u}{\partial x}\right]_j
- \mu \left[\frac{\partial^3 u}{\partial x^3}\right]_j.
\label{eq:rhs_operator}
\end{equation}
The spatial derivatives in equation~\eqref{eq:rhs_operator} are computed
spectrally via equations~\eqref{eq:first_derivative}
and~\eqref{eq:third_derivative}. The nonlinear term
$-\varepsilon u_j(\partial_x u)_j$ introduces quadratic coupling in
Fourier space, which may lead to aliasing errors. However, for the
resolutions employed in the \texttt{sangkuriang} test cases ($N = 512$
or $N = 1024$), these effects remain negligible relative to the specified
tolerances~\cite{ref-19}.

The computational efficiency of the \texttt{sangkuriang} solver is
significantly enhanced through JIT compilation and parallel processing
capabilities provided by Numba~\cite{ref-13}, a dynamic compiler for
Python that translates annotated numerical functions to optimized machine
code via the Low Level Virtual Machine (LLVM) compiler infrastructure. The
pointwise operations in the nonlinear and dispersive terms of
equation~\eqref{eq:rhs_operator} are implemented in specialized functions
decorated with \texttt{@njit(parallel=True, cache=True)}, where the
\texttt{parallel=True} flag enables automatic parallelization of loops
across multiple Central Processing Unit (CPU) cores using Open
Multi-Processing (OpenMP)-style threading~\cite{ref-13}. Specifically, the
elementwise computation of the nonlinear term
$-\varepsilon u_j(\partial_x u)_j$ is structured using \texttt{prange}
(parallel range) rather than Python's standard \texttt{range}, allowing
Numba to distribute loop iterations across available processor threads. The
\texttt{cache=True} option stores compiled machine code to disk,
eliminating recompilation overhead in subsequent runs. The number of
parallel threads is controlled via \texttt{numba.set\_num\_threads(n)},
where $n$ may be specified by the user or defaults to the total number of
available CPU cores detected by the operating system. This parallelization
strategy exploits the embarrassingly parallel nature of pointwise
arithmetic operations on grid data, though the actual performance gain
depends on problem size, memory bandwidth, and hardware
architecture---typical speedups range from $5\times$ to $50\times$
compared to interpreted NumPy operations for grid sizes $N \geq
512$~\cite{ref-13}. The JIT compilation overhead is amortized across
multiple function calls during time integration, making the optimization
particularly effective for the iterative RHS evaluations required by the
adaptive time stepper. It should be noted that spectral derivative
computations via FFT operations remain implemented in NumPy, as these
already employ highly optimized Fastest Fourier Transform in the West
(FFTW) or Intel Math Kernel Library (MKL) libraries that are difficult to
improve upon with custom JIT compilation.

The semi-discrete system~\eqref{eq:semi_discrete} is integrated in time
using the DOP853 method~\cite{ref-21,ref-32}, an eighth-order explicit
Runge--Kutta scheme with embedded fifth and third-order error estimators
for adaptive step size control. This method is accessed through
\texttt{scipy.integrate.solve\_ivp}~\cite{ref-12} with the
\texttt{method='DOP853'} option. The DOP853 scheme requires twelve
function evaluations per accepted step and provides local error estimates
of the form~\cite{ref-22}
\begin{equation}
\text{err}_{\text{local}} =
\|\mathbf{u}_{n+1}^{(8)} - \mathbf{u}_{n+1}^{(5)}\|,
\label{eq:local_error}
\end{equation}
where $\mathbf{u}_{n+1}^{(8)}$ and $\mathbf{u}_{n+1}^{(5)}$ denote the
eighth and fifth-order solution approximations at time $t_{n+1}$,
respectively, and $\|\cdot\|$ represents an appropriately weighted norm.
The step size adaptation follows the standard embedded method
strategy~\cite{ref-22}. If the error estimate~\eqref{eq:local_error}
satisfies
\begin{equation}
\text{err}_{\text{local}} \leq
\text{rtol} \cdot \|\mathbf{u}_{n+1}\| + \text{atol},
\label{eq:error_control}
\end{equation}
where $\text{rtol}$ denotes the relative tolerance and $\text{atol}$ the
absolute tolerance, the step is accepted and the next step size is
computed as
\begin{equation}
\Delta t_{n+1} = 0.9 \, \Delta t_n
\left(\frac{\text{rtol} \cdot \|\mathbf{u}_{n+1}\| + \text{atol}}
{\text{err}_{\text{local}}}\right)^{1/8},
\label{eq:step_size}
\end{equation}
where the exponent $1/8$ corresponds to the order of the method and the
safety factor $0.9$ provides robustness against step rejections. If
condition~\eqref{eq:error_control} is violated, the step is rejected and
recomputed with a reduced step size. The \texttt{sangkuriang}
implementation employs default tolerances of $\text{rtol} = 10^{-10}$ and
$\text{atol} = 10^{-12}$ for test cases 1--3, and tighter tolerances
$\text{rtol} = 10^{-11}$ and $\text{atol} = 10^{-13}$ for the more
complex three-soliton case 4.

The conservation properties described by
equations~\eqref{eq:mass_invariant}--\eqref{eq:energy_invariant} serve as
critical diagnostics for numerical accuracy~\cite{ref-33,ref-8}. In the
oceanographic context, the preservation of these invariants ensures that
the numerical solver does not introduce spurious sources or sinks of wave
energy or momentum that could distort the simulated soliton dynamics. For
the semi-discrete solution
$\mathbf{u}(t) = [u_0(t), u_1(t), \ldots, u_{N-1}(t)]^T$, these
integrals are approximated using the composite trapezoidal
rule~\cite{ref-34}:
\begin{equation}
M(t) \approx \Delta x \sum_{j=0}^{N-1} u_j(t),
\label{eq:discrete_mass}
\end{equation}
\begin{equation}
P(t) \approx \Delta x \sum_{j=0}^{N-1} u_j^2(t),
\label{eq:discrete_momentum}
\end{equation}
\begin{equation}
E(t) \approx \Delta x \sum_{j=0}^{N-1}
\left[\frac{\varepsilon}{2}u_j^3(t)
- \frac{3\mu}{2}\left(\frac{\partial u}{\partial x}\right)_j^2(t)\right],
\label{eq:discrete_energy}
\end{equation}
where the spatial derivative in equation~\eqref{eq:discrete_energy} is
computed spectrally via equation~\eqref{eq:first_derivative}. The
trapezoidal rule achieves second-order accuracy for periodic
functions~\cite{ref-34}, though the spectral accuracy of the solution
itself may allow for higher-order quadrature. The computational
implementation uses \texttt{numpy.trapz}~\cite{ref-11} for numerical
integration. The conservation errors are quantified as maximum relative
deviations over the simulation time span $[0, T_{\text{final}}]$:
\begin{equation}
\text{err}_M = \max_{t \in [0, T_{\text{final}}]}
\frac{|M(t) - M(0)|}{|M(0)|},
\label{eq:mass_error}
\end{equation}
\begin{equation}
\text{err}_P = \max_{t \in [0, T_{\text{final}}]}
\frac{|P(t) - P(0)|}{|P(0)|},
\label{eq:momentum_error}
\end{equation}
\begin{equation}
\text{err}_E = \max_{t \in [0, T_{\text{final}}]}
\frac{|E(t) - E(0)|}{|E(0)|}.
\label{eq:energy_error}
\end{equation}

These metrics provide quantitative measures of the numerical method's
ability to preserve the underlying geometric structure of the KdV equation.
Conservation errors below $10^{-2}$ are generally considered acceptable
for practical simulations, though the \texttt{sangkuriang} implementation
typically achieves errors in the range $10^{-6}$ to $10^{-4}$ depending
on the test case complexity~\cite{ref-33}.

The complete numerical algorithm proceeds as follows. Given initial
condition $u_0(x) = u(x, t=0)$, we sample at grid points to obtain the
initial state vector
$\mathbf{u}(0) = [u_0(x_0), u_0(x_1), \ldots, u_0(x_{N-1})]^T$. The
wavenumber vector $\{k_n\}_{n=0}^{N-1}$ is precomputed according to
equation~\eqref{eq:wavenumber} and stored for repeated use in spectral
derivative evaluations. The Numba JIT compiler is initialized with the
specified number of parallel threads via
\texttt{set\_num\_threads(n\_cores)}, enabling multi-core parallelization
for subsequent RHS evaluations. At each time step, the right-hand side
operator~\eqref{eq:rhs_operator} is evaluated by: (i) computing the FFT
of the current state via equation~\eqref{eq:dft}, (ii) multiplying by
$ik_n$ and $(ik_n)^3$ to obtain derivative Fourier coefficients, (iii)
applying the IFFT via equation~\eqref{eq:idft} to recover physical space
derivatives, and (iv) assembling the nonlinear and dispersive terms using
JIT-compiled parallel functions. The DOP853 integrator advances the
solution from $t_n$ to $t_{n+1}$ according to the adaptive strategy
described by equations~\eqref{eq:error_control}--\eqref{eq:step_size}. At
specified output times $\{t_m\}_{m=0}^{M}$, the solution state is stored
and conservation
laws~\eqref{eq:discrete_mass}--\eqref{eq:discrete_energy} are evaluated
for diagnostic purposes. This process continues until the final time
$T_{\text{final}}$ is reached. The \texttt{sangkuriang} numerical
implementation leverages established scientific computing libraries widely
used in the oceanographic and geoscience communities: NumPy~\cite{ref-11}
for array operations and FFT routines, SciPy~\cite{ref-12} for adaptive
ODE integration, Numba~\cite{ref-13} for JIT compilation and parallel
processing optimization, Matplotlib~\cite{ref-35} for visualization, and
netCDF4~\cite{ref-23} for scientific data output following CF-1.8
conventions. This combination of spectral methods, high-order adaptive
time integration, and parallel computing optimizations may provide both
computational efficiency and numerical accuracy appropriate for
research-grade simulations of idealized oceanic soliton dynamics.

\subsection{Numerical Experiments}

The \texttt{sangkuriang} solver implementation is validated and
demonstrated through four progressively complex test cases that probe
different aspects of KdV soliton dynamics relevant to idealized ocean wave
propagation. These scenarios are designed to assess: (i) baseline
propagation characteristics of isolated solitary waves, as observed when a
single internal soliton traverses a continental shelf~\cite{ref-46};
(ii) interaction dynamics between equal-amplitude waves, analogous to
tidally generated soliton trains with similar amplitudes~\cite{ref-49};
(iii) overtaking collision phenomena arising from amplitude-dependent
velocities, a common occurrence when faster, larger-amplitude internal
waves catch up with smaller predecessors in a wave
packet~\cite{ref-47,ref-51}; and (iv) complex multi-body interactions in
three-soliton systems, representing the rich dynamics of rank-ordered
internal soliton trains frequently observed in satellite imagery of
marginal seas~\cite{ref-48}. While the \texttt{sangkuriang} framework is
general and may be applied to arbitrary initial conditions satisfying the
periodicity constraint, these canonical test cases provide standardized
benchmarks that have been extensively studied in the
literature~\cite{ref-3,ref-6,ref-1} and serve to verify the
implementation's ability to capture the essential physics of nonlinear
dispersive wave propagation in shallow water environments.

The first test case examines the propagation of a single soliton in
isolation, serving as a baseline reference for subsequent multi-soliton
scenarios. This configuration represents the simplest idealization of a
solitary internal wave propagating across a flat-bottomed shelf region.
The initial condition is specified using the exact single-soliton solution
profile~\eqref{eq:soliton_solution} in the form
\begin{equation}
u(x, 0) = \frac{A}{\cosh^2\left[\frac{x - x_0}{w}\right]},
\label{eq:case1_ic}
\end{equation}
with amplitude $A = 4.0$~m, width parameter $w = 2.0$~m, and initial
position $x_0 = -10.0$~m. The computational domain spans
$x \in [-30, 30]$~m discretized with $N = 512$ grid points, providing
spatial resolution $\Delta x \approx 0.117$~m. The physical parameters
are set to $\mu = 0.1$~m$^3$/s and $\varepsilon = 0.2$~m$^{-1}$,
yielding theoretical propagation velocity $v = \varepsilon A/3 \approx
0.267$~m/s according to equation~\eqref{eq:soliton_velocity}. The
simulation extends to final time $T_{\text{final}} = 50.0$~s, during
which the soliton should traverse approximately $13.3$~m while
maintaining its shape. This test case validates the \texttt{sangkuriang}
solver's ability to preserve soliton structure over extended propagation
distances and provides a baseline for assessing conservation law
preservation. The expected relative conservation errors for mass,
momentum, and energy should remain below $10^{-6}$ for this smooth,
non-interacting solution~\cite{ref-33}.

The second test case investigates the interaction between two solitons of
equal amplitude, which may exhibit phase shift phenomena without energy
exchange. This scenario is motivated by observations of tidally generated
internal wave packets in which successive solitons emerge with comparable
amplitudes from a generation site such as a sill or shelf
break~\cite{ref-49,ref-51}. The initial condition comprises a linear
superposition of two sech$^2$ profiles:
\begin{equation}
u(x, 0) = \frac{A_1}{\cosh^2\left[\frac{x - x_1}{w_1}\right]}
+ \frac{A_2}{\cosh^2\left[\frac{x - x_2}{w_2}\right]},
\label{eq:case2_ic}
\end{equation}
with identical amplitudes $A_1 = A_2 = 3.0$~m, widths
$w_1 = w_2 = 2.0$~m, and symmetrically positioned at $x_1 = -15.0$~m
and $x_2 = 15.0$~m. The domain is extended to $x \in [-40, 40]$~m with
$N = 512$ points ($\Delta x \approx 0.156$~m), and the simulation time
increased to $T_{\text{final}} = 60.0$~s. Since both solitons possess
equal amplitudes, they propagate with identical velocities $v = 0.2$~m/s
and thus maintain constant separation. However, the initial linear
superposition~\eqref{eq:case2_ic} is not an exact solution of the
nonlinear KdV equation, necessitating an adjustment phase during which
the wave profiles redistribute energy into soliton modes and dispersive
radiation~\cite{ref-6}. This test case assesses the \texttt{sangkuriang}
solver's handling of multi-soliton configurations and the emergence of
approximate two-soliton solutions from non-exact initial data. The
symmetric configuration also serves as a diagnostic for numerical
artifacts, as any asymmetry in the computed solution would indicate
spatial discretization errors or inadequate time integration accuracy.

The third test case examines the overtaking collision between solitons of
unequal amplitudes, demonstrating the remarkable elastic scattering
property that defines soliton behavior. In the ocean, such overtaking
events occur routinely when a train of rank-ordered internal solitary
waves propagates across a shelf, with the leading (largest) soliton
moving fastest according to the amplitude--velocity
relation~\eqref{eq:soliton_velocity}, and occasionally overtaking smaller
waves that were generated earlier or at a different
source~\cite{ref-47,ref-48}. The initial condition takes the form of
equation~\eqref{eq:case2_ic} with asymmetric parameters: $A_1 = 6.0$~m,
$w_1 = 1.5$~m, $x_1 = -18.0$~m for the leading (taller) soliton, and
$A_2 = 2.0$~m, $w_2 = 2.5$~m, $x_2 = -5.0$~m for the trailing
(shorter) soliton. According to the velocity
relation~\eqref{eq:soliton_velocity}, the taller soliton propagates at
$v_1 = 0.4$~m/s while the shorter moves at $v_2 \approx 0.133$~m/s,
ensuring that the faster wave overtakes the slower one during the
simulation. The domain spans $x \in [-40, 40]$~m with $N = 512$ points,
and the final time is extended to $T_{\text{final}} = 70.0$~s to observe
the complete collision sequence and subsequent separation. This scenario
probes the nonlinear interaction dynamics most severely, as the collision
involves significant amplitude modulation and temporary waveform
distortion~\cite{ref-3}. The remarkable theoretical prediction---that both
solitons emerge from the collision with their original amplitudes and
velocities intact, differing only by phase shifts---provides a stringent
test of the numerical method's ability to capture the integrable structure
of the KdV equation~\cite{ref-5}. Conservation errors may increase during
the collision phase due to enhanced nonlinearity, but should remain below
$10^{-4}$ for the employed tolerances.

The fourth test case extends the complexity to a three-soliton system,
creating a choreography of multiple interactions that tests the
\texttt{sangkuriang} solver's robustness under more challenging
conditions. This configuration serves as an idealized representation of
the rank-ordered internal soliton packets widely documented in the South
China Sea, the Strait of Gibraltar, and other energetic oceanic
environments, where three or more solitary waves of decreasing amplitude
propagate as a coherent group~\cite{ref-46,ref-49}. The initial condition
generalizes equation~\eqref{eq:case2_ic} to three components:
\begin{equation}
u(x, 0) = \sum_{j=1}^{3}
\frac{A_j}{\cosh^2\left[\frac{x - x_j}{w_j}\right]},
\label{eq:case4_ic}
\end{equation}
with amplitudes $A_1 = 7.0$~m, $A_2 = 4.0$~m, $A_3 = 2.5$~m, widths
$w_1 = 1.2$~m, $w_2 = 1.8$~m, $w_3 = 2.2$~m, and initial positions
$x_1 = -25.0$~m, $x_2 = -10.0$~m, $x_3 = 5.0$~m, respectively. The
corresponding velocities are $v_1 \approx 0.467$~m/s, $v_2 \approx
0.267$~m/s, and $v_3 \approx 0.167$~m/s, ordered such that each faster
soliton will sequentially overtake the slower ones ahead. To accommodate
the increased spatial extent of the three-wave system and provide adequate
resolution for the narrowest soliton, the domain is enlarged to
$x \in [-50, 50]$~m with enhanced resolution $N = 1024$
($\Delta x \approx 0.098$~m). The simulation extends to
$T_{\text{final}} = 80.0$~s with tightened tolerances
$\text{rtol} = 10^{-11}$ and $\text{atol} = 10^{-13}$ to maintain
accuracy through multiple collision events. This test case may exhibit
complex interaction patterns including transient amplitude amplifications
during collisions, the generation of small-amplitude dispersive radiation,
and intricate phase shift accumulations~\cite{ref-6,ref-1}. The
successful navigation of these multiple interactions while preserving
conservation laws and soliton identities demonstrates the
\texttt{sangkuriang} solver's capability for research-grade simulations of
multi-soliton dynamics relevant to oceanic internal wave studies.

The physical parameters $\mu = 0.1$~m$^3$/s and
$\varepsilon = 0.2$~m$^{-1}$ are held constant across all test cases to
facilitate direct comparison of dynamical behaviors. These values are
representative of the weakly nonlinear, weakly dispersive regime
characteristic of long gravity waves in shallow coastal
waters~\cite{ref-26}, and ensure that the balance between nonlinearity
and dispersion falls within the regime where soliton solutions are
well-defined. The specific choices of soliton amplitudes and positions in
each test case are informed by established benchmarks in the soliton
literature~\cite{ref-3,ref-20} and are designed to produce qualitatively
distinct interaction scenarios within computationally tractable simulation
times. While these four cases provide comprehensive validation of the
\texttt{sangkuriang} solver's core functionality, the implementation
supports arbitrary initial conditions through the \texttt{SechProfile},
\texttt{TanhProfile}, \texttt{GaussianProfile}, and \texttt{MultiSoliton}
classes, enabling users to investigate alternative scenarios such as
soliton generation from non-sech$^2$ initial data, wave packet
decomposition, or soliton interactions with continuous wave
backgrounds---all of which are relevant to understanding observed oceanic
internal wave phenomena~\cite{ref-50,ref-51}.

The output from each \texttt{sangkuriang} simulation includes the complete
spatiotemporal field $u(x,t)$, conservation law time series $M(t)$,
$P(t)$, $E(t)$ as defined in
equations~\eqref{eq:discrete_mass}--\eqref{eq:discrete_energy}, relative
conservation errors~\eqref{eq:mass_error}--\eqref{eq:energy_error}, and
diagnostic information regarding computational cost (number of function
evaluations, wall clock time). Data are stored in netCDF4
format~\cite{ref-23} following CF-1.8 metadata conventions, facilitating
interoperability with standard oceanographic data analysis tools and
ensuring compatibility with widely used geoscience software packages.
Animated visualizations are generated showing the evolving wave profile
in three-dimensional space-time representations, with conservation law
diagnostics overlaid to provide immediate visual feedback on solution
quality. These outputs enable both qualitative assessment of soliton
dynamics and quantitative verification of the numerical implementation's
fidelity to the underlying physics encoded in the KdV
equation~\eqref{eq:kdv_standard}.

\subsection{Data Analysis}

The diagnostic analysis of simulation outputs is performed independently
of the \texttt{sangkuriang} solver library through a dedicated suite of
post-processing routines implemented in Python. These routines employ
NumPy~\cite{ref-11} for array operations and FFT, SciPy~\cite{ref-12} for
signal processing and distance computations, and Matplotlib~\cite{ref-35}
for visualization. Simulation data archived in netCDF4
format~\cite{ref-23} following CF-1.8 conventions are subjected to four
complementary analyses: verification of conservation law preservation,
soliton trajectory tracking with velocity validation, spectral
information-theoretic characterization, and phase space recurrence
quantification. Each analysis extracts quantitative metrics that
collectively assess both numerical accuracy and physical fidelity of the
computed solutions. Together, these diagnostics provide a comprehensive
assessment framework applicable not only to the idealized KdV system
studied here but also to more complex ocean wave models where analytical
benchmarks may be unavailable~\cite{ref-50,ref-51}.

The conservation analysis quantifies the temporal stability of the three
KdV invariants defined in
equations~\eqref{eq:discrete_mass}--\eqref{eq:discrete_energy}. Let
$Q(t)$ denote any of the conserved quantities $M(t)$, $P(t)$, or $E(t)$
evaluated at discrete output times $\{t_n\}_{n=0}^{N_t-1}$, where $N_t$
is the total number of stored time snapshots. The instantaneous relative
deviation from the initial state is defined as
\begin{equation}
\delta_Q(t_n) = \frac{Q(t_n) - Q(t_0)}{|Q(t_0)|},
\label{eq:relative_deviation}
\end{equation}
which measures the fractional departure of $Q$ from its initial value
$Q(t_0) = Q_0$. Three statistical measures characterize the behavior of
$\delta_Q$ over the simulation interval. The maximum relative error,
\begin{equation}
\text{err}_Q = \max_{0 \leq n < N_t} |\delta_Q(t_n)|,
\label{eq:max_relative_error}
\end{equation}
captures the worst-case deviation and is stored directly in the netCDF
output files. The RMS error,
\begin{equation}
\sigma_Q = \left[\frac{1}{N_t}\sum_{n=0}^{N_t-1}
\delta_Q^2(t_n)\right]^{1/2},
\label{eq:rms_conservation_error}
\end{equation}
provides a time-averaged measure of conservation fidelity that is less
sensitive to isolated outliers. To distinguish systematic drift from
bounded oscillations---a distinction important for assessing whether
numerical errors would accumulate to unphysical levels over the long
propagation distances typical of oceanic solitary waves---the linear drift
rate $\gamma_Q$ is extracted via ordinary least-squares regression.
Defining the design matrix $\mathbf{A} \in \mathbb{R}^{N_t \times 2}$
with elements $A_{n1} = t_n$ and $A_{n2} = 1$, and the observation vector
$\mathbf{b} = [\delta_Q(t_0), \delta_Q(t_1), \ldots,
\delta_Q(t_{N_t-1})]^T$, the drift rate is obtained as the first
component of the least-squares solution
\begin{equation}
[\gamma_Q, \beta_Q]^T =
(\mathbf{A}^T\mathbf{A})^{-1}\mathbf{A}^T\mathbf{b},
\label{eq:drift_regression}
\end{equation}
where $\beta_Q$ is an intercept term. A nonzero $\gamma_Q$ [s$^{-1}$]
indicates secular accumulation of numerical error, whereas $\gamma_Q
\approx 0$ with finite $\sigma_Q$ suggests oscillatory but bounded
deviations consistent with symplectic or near-symplectic integration.

Soliton trajectory extraction proceeds through systematic identification
and tracking of local amplitude maxima in the spatiotemporal field
$u(x_j, t_n)$. This procedure is analogous to the peak-tracking methods
employed in observational oceanography to identify and follow individual
internal solitary waves in mooring time series or satellite image
sequences~\cite{ref-47,ref-49}. At each output time $t_n$, the discrete
spatial profile $\{u_j^{(n)}\}_{j=0}^{N-1} \equiv \{u(x_j,
t_n)\}_{j=0}^{N-1}$ is analyzed using the
\texttt{scipy.signal.find\_peaks} routine~\cite{ref-12}, which identifies
indices $j^*$ satisfying $u_{j^*}^{(n)} > u_{j^*-1}^{(n)}$,
$u_{j^*}^{(n)} > u_{j^*+1}^{(n)}$, subject to additional constraints.
Two filtering parameters govern peak selection: a height threshold
$u_{\text{th}}$ requiring $u_{j^*}^{(n)} > u_{\text{th}}$, and a minimum
separation distance $d_{\text{min}}$ (in grid points) enforcing
$|j_1^* - j_2^*| > d_{\text{min}}$ between distinct peaks. For the test
cases examined, $u_{\text{th}} = 0.5$~m for multi-soliton configurations
and $u_{\text{th}} = 1.0$~m for single-soliton runs, with
$d_{\text{min}} = 10$ grid points throughout. Each detected peak yields a
triplet $(t_n, x_{j^*}, u_{j^*}^{(n)})$ recording the time, spatial
position, and amplitude.

For configurations containing $K$ solitons, individual trajectories must
be disambiguated from the aggregate peak data. This is accomplished
through amplitude-based sorting at each time instant: letting
$\{(x_k^{(n)}, A_k^{(n)})\}_{k=1}^{K_n}$ denote the $K_n$ peaks
detected at time $t_n$ with positions $x_k^{(n)}$ and amplitudes
$A_k^{(n)}$, these are reordered such that $A_1^{(n)} \geq A_2^{(n)}
\geq \cdots \geq A_{K_n}^{(n)}$. The $k$-th soliton track is then
defined as the sequence $\{(t_n, x_k^{(n)})\}$ across all times where at
least $k$ peaks were detected. This ranking scheme exploits the
amplitude-velocity ordering inherent to KdV solitons: since taller
solitons propagate faster according to
equation~\eqref{eq:soliton_velocity}, amplitude rank provides a consistent
identifier even through collision events where spatial ordering temporarily
inverts. The same amplitude-rank ordering is a well-known feature of
oceanic internal solitary wave packets, where the leading wave is
typically the largest~\cite{ref-46,ref-48}.

The propagation velocity of each tracked soliton is determined by linear
regression of position against time. For the $k$-th soliton trajectory
restricted to a temporal window $[t_a, t_b]$ containing $N_k$ data
points, the measured velocity $v_k^{\text{meas}}$ minimizes the sum of
squared residuals:
\begin{equation}
v_k^{\text{meas}} = \underset{v}{\arg\min}
\sum_{n: t_n \in [t_a, t_b]}
\left[x_k^{(n)} - v t_n - x_k^{(0)}\right]^2.
\label{eq:velocity_lsq}
\end{equation}
The explicit solution via \texttt{numpy.polyfit}~\cite{ref-11} yields
$v_k^{\text{meas}}$ as the slope of the fitted line
$x_k(t) = v_k^{\text{meas}} t + x_k^{(0)}$. Goodness-of-fit is
quantified by the coefficient of determination
\begin{equation}
R_k^2 = 1 - \frac{\sum_n (x_k^{(n)} - v_k^{\text{meas}} t_n
- x_k^{(0)})^2}{\sum_n (x_k^{(n)} - \bar{x}_k)^2},
\label{eq:r_squared}
\end{equation}
where $\bar{x}_k = N_k^{-1}\sum_n x_k^{(n)}$ is the mean position.
Values $R_k^2 \to 1$ indicate linear trajectories consistent with
constant-velocity propagation. The mean soliton amplitude over the fitting
interval,
\begin{equation}
\bar{A}_k = \frac{1}{N_k}\sum_{n: t_n \in [t_a, t_b]} A_k^{(n)},
\label{eq:mean_amplitude}
\end{equation}
together with its standard deviation $\sigma_{A_k}$, enables comparison
against the theoretical velocity
relation~\eqref{eq:soliton_velocity}. The theoretical prediction
$v_k^{\text{theo}} = \varepsilon \bar{A}_k / 3$ should match
$v_k^{\text{meas}}$ for well-resolved soliton solutions, providing
validation of both numerical accuracy and the soliton character of
computed waveforms.

Spectral analysis employs the DFT~\eqref{eq:dft} to characterize the
wavenumber content of each spatial snapshot. For the field
$\{u_j^{(n)}\}_{j=0}^{N-1}$ at time $t_n$, the power spectral density is
computed as
\begin{equation}
P_m^{(n)} = \frac{1}{N}|\hat{u}_m^{(n)}|^2,
\quad m = 0, 1, \ldots, N-1,
\label{eq:power_spectrum}
\end{equation}
where $\hat{u}_m^{(n)}$ denotes the $m$-th Fourier coefficient obtained
via \texttt{numpy.fft.fft}~\cite{ref-11}. The corresponding wavenumbers
$k_m$ follow from equation~\eqref{eq:wavenumber} and are generated using
\texttt{numpy.fft.fftfreq}. Normalization of the power spectrum yields a
discrete probability distribution over wavenumber space:
\begin{equation}
p_m^{(n)} = \frac{P_m^{(n)}}{\sum_{\ell=0}^{N-1} P_\ell^{(n)}},
\quad \sum_{m=0}^{N-1} p_m^{(n)} = 1,
\label{eq:spectral_probability}
\end{equation}
representing the fractional energy content at wavenumber $k_m$. This
spectral probability distribution forms the basis for three
information-theoretic measures that characterize distinct aspects of the
wave field structure.

The spectral entropy~\cite{ref-36} quantifies the breadth of spectral
energy distribution through the normalized Shannon entropy of
$\{p_m^{(n)}\}$:
\begin{equation}
S_k^{(n)} = -\frac{1}{\ln N}\sum_{m=0}^{N-1}
p_m^{(n)} \ln p_m^{(n)},
\label{eq:spectral_entropy}
\end{equation}
where terms with $p_m^{(n)} < 10^{-15}$ are excluded from the summation
to avoid numerical singularities. The normalization by $\ln N$ ensures
$S_k^{(n)} \in [0, 1]$, with $S_k^{(n)} = 0$ corresponding to a
monochromatic signal (all energy concentrated at a single wavenumber) and
$S_k^{(n)} = 1$ to uniform spectral distribution (white noise). Soliton
profiles, characterized by smooth $\text{sech}^2$ envelopes with
exponentially decaying Fourier spectra, exhibit intermediate entropy
values that remain approximately constant during propagation. Changes in
spectral entropy during soliton interactions provide a measure of how the
wave field's spectral complexity evolves---information that is relevant to
understanding energy cascades in oceanic internal wave
fields~\cite{ref-51}.

The L\'{o}pez--Ruiz--Mancini--Calbet (LMC) statistical
complexity~\cite{ref-37} captures the interplay between disorder and
departure from equilibrium through the product
\begin{equation}
C_{\text{LMC}}^{(n)} = H^{(n)} \times D^{(n)},
\label{eq:lmc_complexity}
\end{equation}
where $H^{(n)} = S_k^{(n)}$ is the normalized spectral entropy defined in
equation~\eqref{eq:spectral_entropy}, and the disequilibrium
\begin{equation}
D^{(n)} = \sum_{m=0}^{N-1}
\left(p_m^{(n)} - \frac{1}{N}\right)^2
\label{eq:disequilibrium}
\end{equation}
measures the squared Euclidean distance between the observed distribution
$\{p_m^{(n)}\}$ and the uniform distribution $\{1/N\}$. This construction
ensures $C_{\text{LMC}}^{(n)} = 0$ for both perfectly ordered states
($H^{(n)} = 0$, all energy at one mode) and maximally disordered states
($D^{(n)} = 0$, uniform distribution), attaining nonzero values only for
intermediate configurations exhibiting nontrivial
structure~\cite{ref-37}. The LMC complexity thus distinguishes structured
nonequilibrium states from both trivial order and featureless randomness,
offering a single scalar diagnostic for the degree of organized wave
structure in the computed solutions.

The Fisher information~\cite{ref-38,ref-39,ref-40} provides a
complementary characterization of spatial localization by measuring the
``sharpness'' of the amplitude distribution. Defining the normalized
spatial probability density
\begin{equation}
\rho_j^{(n)} = \frac{|u_j^{(n)}| + \epsilon_{\text{reg}}}
{\sum_{\ell=0}^{N-1} (|u_\ell^{(n)}|
+ \epsilon_{\text{reg}}) \Delta x},
\label{eq:spatial_density}
\end{equation}
where $\epsilon_{\text{reg}} = 10^{-15}$ is a regularization parameter
preventing division by zero, the discrete Fisher information is computed
as
\begin{equation}
F^{(n)} = \sum_{j=0}^{N-1} \frac{1}{\rho_j^{(n)}}
\left(\frac{\partial \rho}{\partial x}\bigg|_j^{(n)}\right)^2 \Delta x.
\label{eq:fisher_discrete}
\end{equation}
The spatial derivative $(\partial_x \rho)_j^{(n)}$ is evaluated using
second-order central differences via
\texttt{numpy.gradient}~\cite{ref-11}:
\begin{equation}
\frac{\partial \rho}{\partial x}\bigg|_j^{(n)}
\approx \frac{\rho_{j+1}^{(n)} - \rho_{j-1}^{(n)}}{2\Delta x},
\label{eq:central_difference}
\end{equation}
with one-sided differences at domain boundaries. Higher Fisher information
indicates sharper spatial gradients and hence more localized wave
structures; for soliton solutions, $F^{(n)}$ scales inversely with the
characteristic width parameter $w$ appearing in
equation~\eqref{eq:soliton_solution}. During soliton collisions, transient
increases in $F^{(n)}$ reflect the temporary steepening of gradients as
wave profiles overlap, providing a quantitative signature of the
interaction events that parallels the steepening observed in field
measurements of colliding internal solitary waves~\cite{ref-47}.

Phase space analysis employs recurrence quantification analysis
(RQA)~\cite{ref-41,ref-42} to characterize the dynamical structure of
simulation trajectories. For integrable Hamiltonian systems possessing
multiple conservation laws, trajectories are confined to low-dimensional
invariant manifolds in the full phase space; deviations from this
structure would indicate either numerical error or departure from
integrability. The analysis proceeds by embedding the temporal evolution
in a three-dimensional conservation space with coordinates
\begin{equation}
\mathbf{y}^{(n)} = \left(\frac{M(t_n)}{M_0}, \frac{P(t_n)}{P_0},
\frac{E(t_n)}{E_0}\right)^T \in \mathbb{R}^3,
\label{eq:conservation_embedding}
\end{equation}
where normalization by initial values $M_0$, $P_0$, $E_0$ renders the
components dimensionless and of comparable magnitude. For a perfectly
conservative numerical scheme, $\mathbf{y}^{(n)} = (1, 1, 1)^T$ for all
$n$; deviations from this ideal point trace out a trajectory reflecting
cumulative numerical errors.

The pairwise Euclidean distance matrix
$\mathbf{D} \in \mathbb{R}^{N_t \times N_t}$ with elements
\begin{equation}
D_{ij} = \|\mathbf{y}^{(i)} - \mathbf{y}^{(j)}\|
= \left[\sum_{\alpha=1}^{3}
(y_\alpha^{(i)} - y_\alpha^{(j)})^2\right]^{1/2}
\label{eq:distance_matrix}
\end{equation}
is computed efficiently using
\texttt{scipy.spatial.distance.pdist} and
\texttt{squareform}~\cite{ref-12}. A recurrence threshold $\varepsilon$
is determined adaptively as the $q$-th percentile of the nonzero elements
of $\mathbf{D}$, with $q = 10$--$15$ providing robust results across test
cases. The binary recurrence matrix~\cite{ref-41} is then constructed as
\begin{equation}
R_{ij} = \Theta(\varepsilon - D_{ij}) =
\begin{cases}
1, & D_{ij} < \varepsilon, \\
0, & D_{ij} \geq \varepsilon,
\end{cases}
\label{eq:recurrence_matrix}
\end{equation}
where $\Theta(\cdot)$ denotes the Heaviside step function. By
construction, $R_{ii} = 1$ for all $i$ (self-recurrence), and
$R_{ij} = R_{ji}$ (symmetry).

Two scalar metrics extracted from the recurrence matrix characterize the
dynamical behavior. The recurrence rate
\begin{equation}
\text{RR} = \frac{1}{N_t(N_t - 1)}
\sum_{\substack{i,j=0 \\ i \neq j}}^{N_t-1} R_{ij}
\label{eq:recurrence_rate}
\end{equation}
quantifies the density of recurrent pairs, excluding the trivial diagonal
$i = j$. Higher RR indicates that the trajectory repeatedly visits
similar regions of conservation space, consistent with bounded oscillatory
errors rather than secular drift. The determinism
\begin{equation}
\text{DET} = \frac{\sum_{\ell \geq \ell_{\min}}
\ell \cdot P(\ell)}{\sum_{i \neq j} R_{ij}}
\label{eq:determinism}
\end{equation}
measures the fraction of recurrence points forming diagonal line
structures of length at least $\ell_{\min}$ (typically $\ell_{\min} = 2$).
Here $P(\ell)$ denotes the number of diagonal lines of exactly length
$\ell$ in the recurrence matrix, where a diagonal line at offset $k > 0$
consists of consecutive entries
$R_{i,i+k} = R_{i+1,i+1+k} = \cdots = 1$. Diagonal structures arise when
segments of the trajectory evolve in parallel, a hallmark of deterministic
dynamics on regular (non-chaotic) attractors~\cite{ref-42}. For integrable
systems such as the KdV equation, DET $\to 1$ reflects confinement to
invariant tori, whereas chaotic systems exhibit DET $\ll 1$ due to
sensitive dependence on initial conditions disrupting parallel trajectory
segments. In the context of ocean wave dynamics, high determinism values
provide confidence that the numerical solver faithfully preserves the
regular dynamical structure expected of the idealized KdV system, which is
a prerequisite for reliable long-time simulations of solitary wave
propagation and interaction~\cite{ref-6}.

Complementary phase space characterization is obtained through analysis of
Fourier mode amplitudes. The leading $K$ complex Fourier coefficients
$\{\hat{u}_m^{(n)}\}_{m=0}^{K-1}$ (typically $K = 3$--$5$) extracted
from each temporal snapshot define a trajectory in a $2K$-dimensional
space (accounting for real and imaginary parts). Projection onto the real
components of the first two modes,
$(\text{Re}[\hat{u}_0^{(n)}], \text{Re}[\hat{u}_1^{(n)}])$, yields a
two-dimensional phase portrait that visualizes the modal dynamics. For
the KdV equation, the zeroth mode
$\hat{u}_0^{(n)} = N^{-1}\sum_j u_j^{(n)}$ is proportional to the
spatial mean (and hence to the mass $M$), while higher modes capture
spatial structure. Bounded, closed orbits in this projection confirm the
quasi-periodic character expected of integrable dynamics, whereas
space-filling or divergent trajectories would signal chaotic or unstable
behavior.

\section{Results}

The \texttt{sangkuriang} solver was validated through four test cases of
increasing complexity, executed on a laptop-class workstation (Lenovo
ThinkPad P52s with Intel Core i7-8550U processor at 4.0~GHz, 8 logical
cores) running Fedora Linux 39 (kernel 6.11.9-100.fc39.x86\_64). This
hardware configuration represents a modest computational platform typical
of those available to individual researchers in oceanography and geoscience
departments, permitting assessment of the solver's practical utility for
idealized ocean wave studies without access to high-performance computing
resources.

Table~\ref{tab:computational_performance} summarizes the computational
cost for each test case. The single-soliton simulation (Case~1) required
208,202 adaptive time steps to integrate over the 50~s interval,
completing in approximately 152~s of wall-clock time. The adaptive time
stepper selected finer temporal resolution for this case due to the smaller
spatial domain ($L = 60$~m) and correspondingly higher wavenumber content
relative to Cases~2 and~3. The two equal-amplitude soliton simulation
(Case~2) completed in 75.7~s using 98,612 steps, while the overtaking
collision (Case~3) required 115,658 steps and 87.6~s. The three-soliton
configuration (Case~4), employing doubled spatial resolution ($N = 1024$)
and tightened tolerances ($\text{rtol} = 10^{-11}$,
$\text{atol} = 10^{-13}$), demanded 551,858 steps and 534~s of
computation time. These timings suggest that the Numba-accelerated
implementation achieves reasonable throughput for exploratory research into
idealized ocean wave dynamics, though the scaling with grid resolution and
tolerance warrants consideration for production simulations involving
extended ocean domains.

\begin{table}[H]
\centering
\caption{Computational performance metrics for the four test cases.
All simulations employed 8 parallel threads via Numba JIT
compilation.\label{tab:computational_performance}}
\small
\begin{tabular}{lccccc}
\toprule
\textbf{Case} & \textbf{$N$} & \textbf{$\Delta x$ [m]} &
\textbf{$T_{\text{final}}$ [s]} & \textbf{Steps} &
\textbf{CPU Time [s]} \\
\midrule
1 (Single soliton)  & 512  & 0.117 & 50 & 208,202 & 152.0 \\
2 (Two equal)       & 512  & 0.157 & 60 &  98,612 &  75.7 \\
3 (Collision)       & 512  & 0.157 & 70 & 115,658 &  87.6 \\
4 (Three solitons)  & 1024 & 0.098 & 80 & 551,858 & 534.4 \\
\bottomrule
\end{tabular}
\end{table}

The throughput, measured in time steps per second, remained relatively
consistent across Cases~1--3 at approximately 1300--1370 steps/s,
decreasing to roughly 1030 steps/s for Case~4. This reduction likely
reflects the increased memory bandwidth demands associated with the larger
grid and the superlinear scaling of FFT operations, which nominally follow
$\mathcal{O}(N \log N)$ complexity~\cite{ref-30}. Nevertheless, all
simulations completed within times amenable to interactive research
workflows, enabling rapid parameter exploration in idealized ocean soliton
studies.

\begin{figure}[H]
\centering
\includegraphics[width=\textwidth]{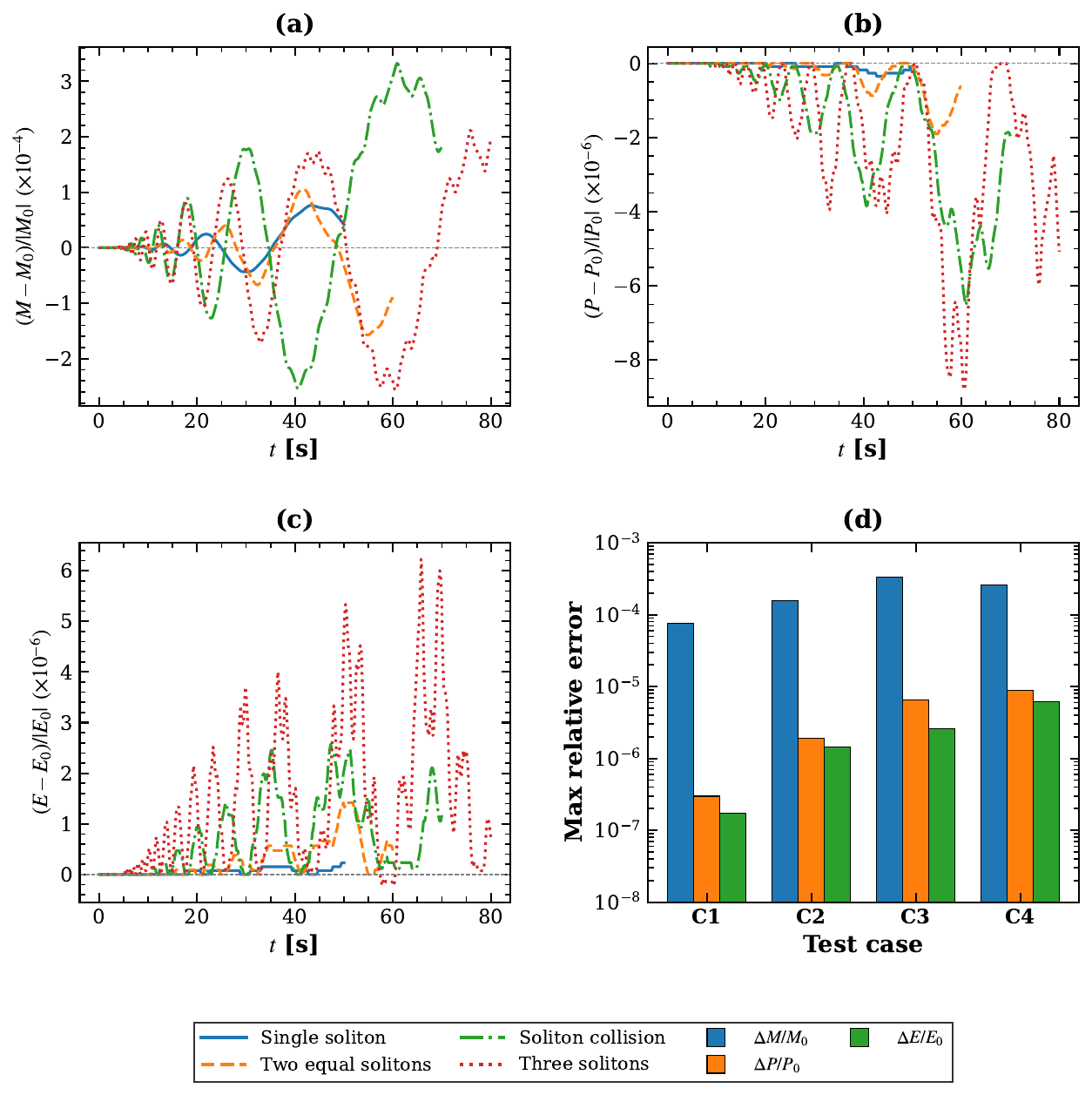}
\caption{Conservation law diagnostics.
(\textbf{a})~Relative mass deviation $(M - M_0)/|M_0|$;
(\textbf{b})~relative momentum deviation $(P - P_0)/|P_0|$;
(\textbf{c})~relative energy deviation $(E - E_0)/|E_0|$;
(\textbf{d})~maximum relative errors across all cases on logarithmic
scale. Solid blue: single soliton (C1); dashed orange: two equal solitons
(C2); dash-dotted green: soliton collision (C3); dotted red: three
solitons (C4).\label{fig:conservation}}
\end{figure}

The preservation of the three KdV invariants---mass $M$, momentum $P$, and
energy $E$ as defined in
equations~\eqref{eq:discrete_mass}--\eqref{eq:discrete_energy}---provides
a stringent test of numerical accuracy. For ocean wave applications, the
faithful conservation of these quantities ensures that the solver does not
introduce spurious wave energy sources or sinks that could contaminate the
simulated dynamics~\cite{ref-33}. Figure~\ref{fig:conservation} displays
the temporal evolution of relative conservation errors for all test cases.

The single-soliton case exhibited the smallest conservation errors, with
maximum relative deviations of $7.64 \times 10^{-5}$ for mass,
$2.98 \times 10^{-7}$ for momentum, and $1.73 \times 10^{-7}$ for energy.
These values are consistent with the specified integration tolerances and
confirm that the pseudo-spectral spatial discretization introduces
negligible additional error for smooth, non-interacting solutions
representative of an isolated oceanic solitary wave. The root-mean-square
(RMS) errors were approximately half the maximum values
($3.55 \times 10^{-5}$, $1.30 \times 10^{-7}$, and $8.28 \times
10^{-8}$, respectively), indicating that the deviations represent bounded
oscillations rather than secular drift. Linear regression of the
conservation time series yielded drift rates on the order of
$10^{-9}$~s$^{-1}$ for momentum and energy, suggesting that the dominant
error contribution arises from the non-symplectic character of the DOP853
integrator rather than from systematic bias~\cite{ref-22}.

Multi-soliton configurations exhibited moderately larger conservation
errors, as anticipated from the increased solution complexity during
interaction events. The two equal-amplitude soliton case (Case~2) showed
maximum errors of $1.57 \times 10^{-4}$ (mass), $1.90 \times 10^{-6}$
(momentum), and $1.45 \times 10^{-6}$ (energy). The overtaking collision
(Case~3) produced the largest mass error at $3.33 \times 10^{-4}$, with
momentum and energy errors of $6.49 \times 10^{-6}$ and $2.60 \times
10^{-6}$, respectively. Figure~\ref{fig:conservation}a--c reveals that
the conservation deviations exhibit transient excursions correlated with
collision events, particularly evident in the three-soliton case where
multiple interactions occur. Such transient increases in conservation error
during wave--wave interaction are expected physically, as the nonlinear
coupling intensifies when soliton profiles overlap~\cite{ref-6}. Despite
the increased complexity of Case~4, the maximum errors
($2.56 \times 10^{-4}$, $8.86 \times 10^{-6}$, $6.20 \times 10^{-6}$)
remained comparable to or smaller than those of Case~3, likely reflecting
the benefits of enhanced spatial resolution and tighter tolerances
employed for this configuration.

The bar chart in Figure~\ref{fig:conservation}d presents a comparative
summary of maximum relative errors across all cases. A consistent
hierarchy emerges: mass conservation errors exceed momentum errors by
approximately two orders of magnitude, which in turn exceed energy errors.
This ordering may reflect the different sensitivities of each integral to
high-wavenumber components of the solution, with the energy
functional~\eqref{eq:energy_invariant} involving spatial derivatives that
weight the well-resolved low-wavenumber content more heavily. Across all
test cases, the conservation errors remain well below the $10^{-2}$
threshold commonly adopted as acceptable for practical simulations of
Hamiltonian systems~\cite{ref-33}, suggesting that the
\texttt{sangkuriang} implementation maintains adequate fidelity to the
underlying physics governing idealized ocean soliton propagation.

\begin{figure}[H]
\centering
\includegraphics[width=\textwidth]{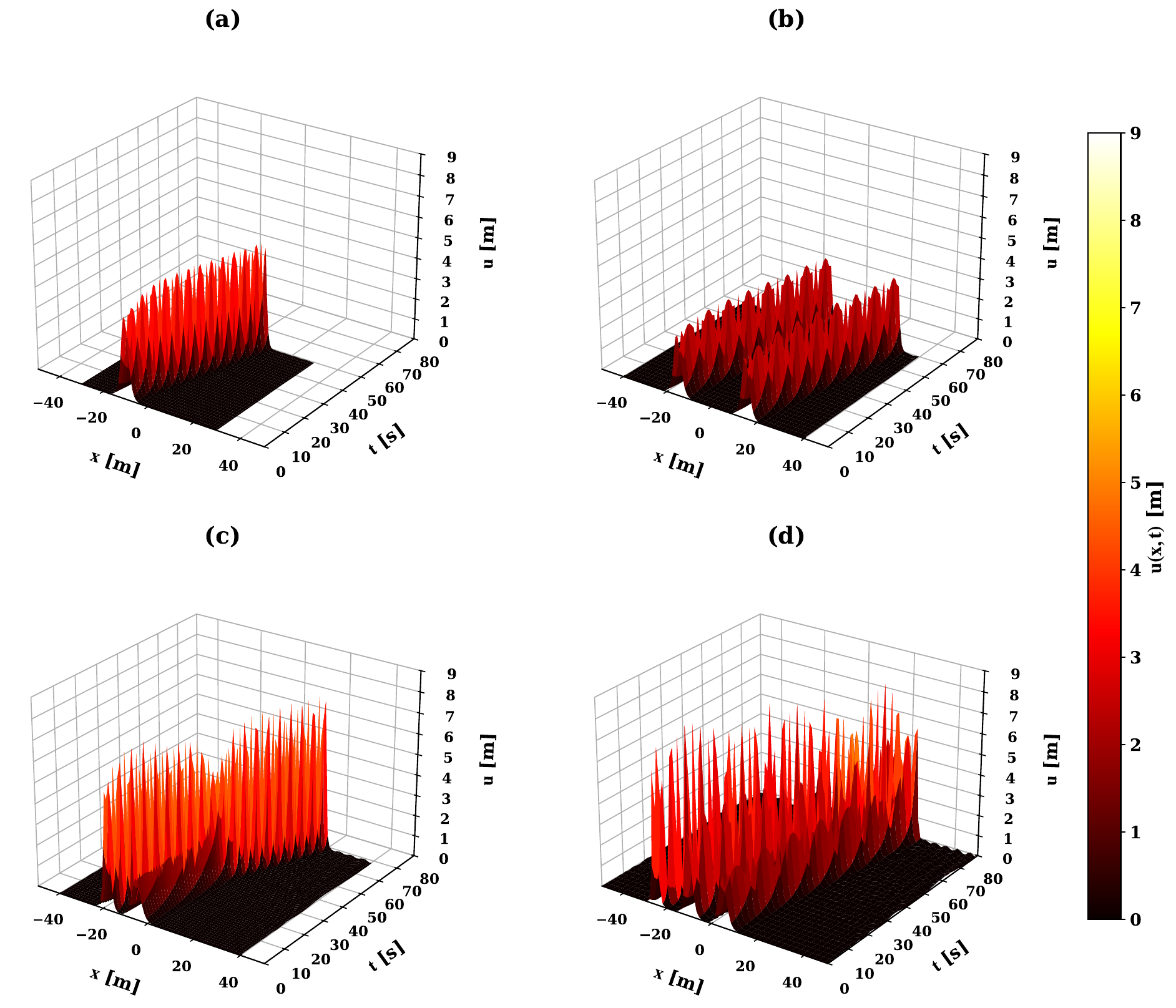}
\caption{Three-dimensional spatiotemporal evolution of the wave amplitude
$u(x,t)$ for the four test cases.
(\textbf{a})~Single soliton propagating rightward;
(\textbf{b})~two equal-amplitude solitons;
(\textbf{c})~overtaking collision between fast (tall) and slow (short)
solitons;
(\textbf{d})~three-soliton interaction.
Color scale indicates amplitude in meters. View angles: elevation
$25^\circ$, azimuth $-55^\circ$.\label{fig:spatiotemporal}}
\end{figure}

Figure~\ref{fig:spatiotemporal} presents three-dimensional visualizations
of the wave field $u(x,t)$ for each test case, rendered with consistent
axis limits ($x \in [-50, 50]$~m, $t \in [0, 80]$~s, $u \in [0, 9]$~m)
to facilitate direct comparison of the idealized oceanic soliton dynamics.

Panel~(a) illustrates the single-soliton case, where the initial
$\text{sech}^2$ profile propagates from left to right while maintaining
its characteristic shape---the hallmark of a solitary wave in a dispersive
medium. The wave amplitude increases slightly from the initial value of
approximately 4.0~m to a final peak of 5.2~m, with a global maximum of
5.21~m observed during the simulation. This amplitude evolution reflects
the adjustment of the initial condition---which approximates but does not
precisely match the exact soliton solution~\eqref{eq:soliton_solution}---toward
a true soliton state through radiation of small-amplitude dispersive
waves~\cite{ref-6}. Such adjustment dynamics are analogous to the fission
process observed when a tidal bore or internal tide evolves into a
rank-ordered train of solitary waves on a continental
shelf~\cite{ref-46,ref-49}. The peak position advanced by 17.8~m over the
50~s integration interval, corresponding to an average velocity of
0.357~m/s.

The two equal-amplitude soliton configuration, shown in panel~(b), exhibits
parallel propagation of two $A = 3.0$~m solitons initially separated by
30~m. Since the KdV soliton velocity is proportional to amplitude
according to equation~\eqref{eq:soliton_velocity}, these waves maintain
constant separation throughout the simulation, akin to the approximately
equal-amplitude internal waves sometimes observed in the trailing portion
of oceanic soliton packets~\cite{ref-48}. The surface plot reveals two
distinct ridges traversing the space-time domain without intersection,
confirming the expected kinematic behavior. Minor amplitude variations
(final peak 3.66~m versus initial 3.0~m) again indicate adjustment toward
exact soliton profiles.

Panel~(c) displays the overtaking collision scenario, which constitutes the
most stringent test of the solver's ability to capture nonlinear
interaction dynamics. The taller soliton ($A_1 = 6.0$~m), propagating
faster than the shorter wave ($A_2 = 2.0$~m), overtakes and passes through
it near $t \approx 35$~s and $x \approx 0$~m. The characteristic
``X''-shaped intersection pattern in the space-time surface demonstrates
the elastic scattering property of KdV solitons: both waves emerge from
the collision with their identities intact, differing only by phase
shifts~\cite{ref-3}. This elastic interaction is the idealized analog of
the overtaking collisions documented in field observations of internal
solitary waves, where larger waves catch and pass smaller ones with
remarkably little energy exchange~\cite{ref-47,ref-51}. The global maximum
amplitude of 7.55~m, exceeding the linear superposition of the two initial
amplitudes, reflects the transient nonlinear amplification during the
collision phase.

The three-soliton interaction in panel~(d) presents the most complex
dynamics, with multiple collision events occurring as the fastest soliton
($A_1 = 7.0$~m) successively overtakes the intermediate ($A_2 = 4.0$~m)
and slowest ($A_3 = 2.5$~m) waves. The space-time surface exhibits a
sequence of intersection patterns, with a global maximum of 8.26~m
observed during the highest-amplitude collision. Despite this complexity,
the three distinct soliton tracks remain identifiable throughout the
simulation, consistent with the complete integrability of the KdV
equation~\cite{ref-4}. This rank-ordered configuration, with the largest
and fastest wave leading, mirrors the structure of oceanic internal soliton
packets routinely observed in satellite synthetic aperture radar
imagery~\cite{ref-46,ref-49}.

\begin{figure}[H]
\centering
\includegraphics[width=\textwidth]{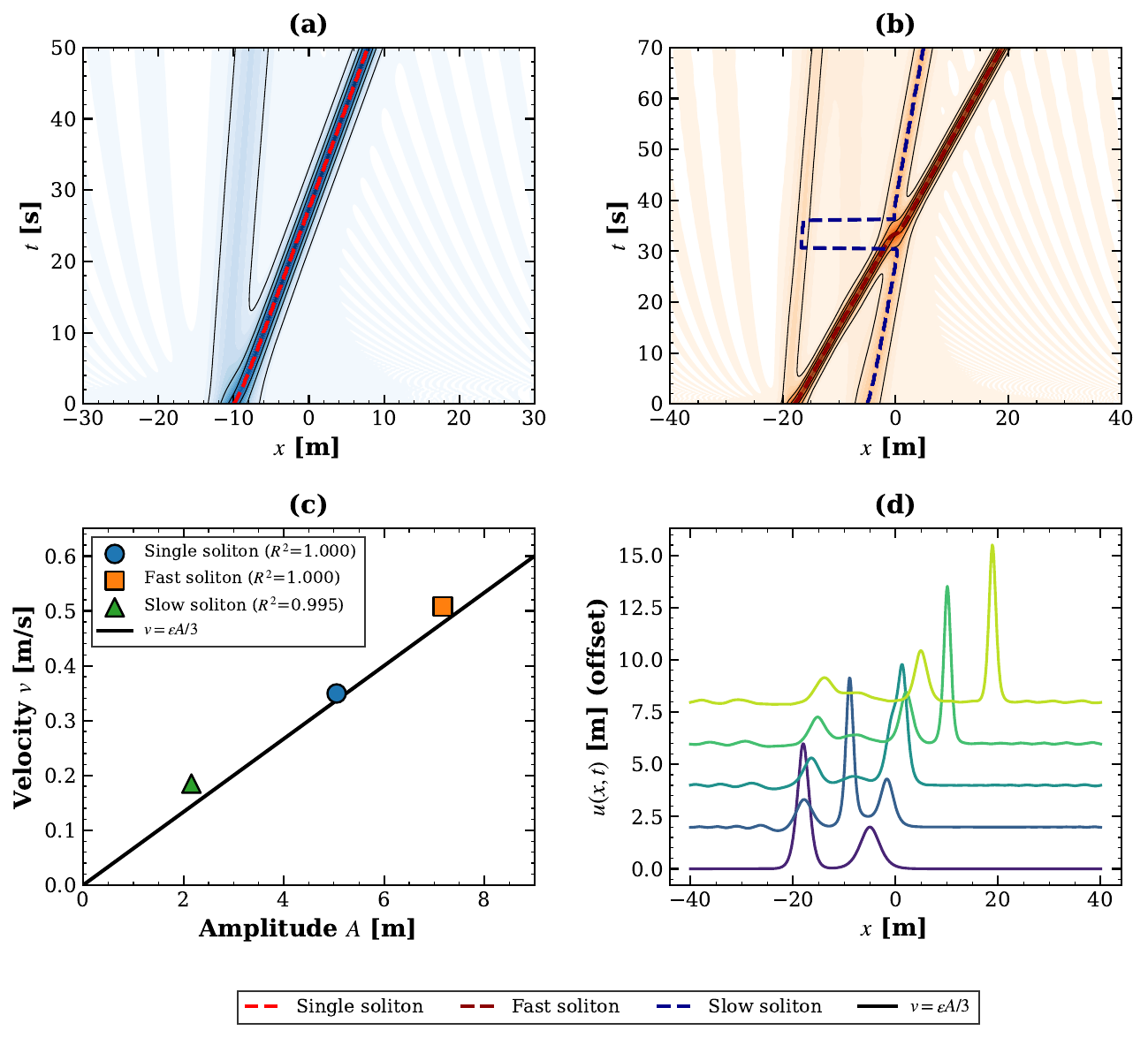}
\caption{Soliton trajectory analysis.
(\textbf{a})~Space-time contour plot for single soliton with tracked
trajectory (dashed line);
(\textbf{b})~collision case showing fast (dark red) and slow (blue)
soliton tracks with collision region indicated;
(\textbf{c})~measured velocity versus amplitude with theoretical relation
$v = \varepsilon A/3$ (solid line);
(\textbf{d})~wave profiles at selected times during collision, vertically
offset for clarity.\label{fig:soliton_dynamics}}
\end{figure}

Quantitative validation of the soliton dynamics requires comparison of
measured propagation velocities against theoretical predictions.
Figure~\ref{fig:soliton_dynamics} presents trajectory data extracted from
the spatiotemporal fields using peak detection and tracking algorithms.

Panel~(a) displays the single-soliton trajectory as a space-time contour
plot with the tracked peak position overlaid. The trajectory exhibits
excellent linearity, with a coefficient of determination $R^2 = 0.9998$
for the linear fit. The measured amplitude of $5.05 \pm 0.29$~m
(mean $\pm$ standard deviation over the simulation) yields a theoretical
velocity of $v_{\text{theo}} = \varepsilon A / 3 = 0.337$~m/s according
to equation~\eqref{eq:soliton_velocity}. The measured velocity of
0.349~m/s exceeds this prediction by approximately 3.6\%, a discrepancy
that may be attributed to the amplitude evolution during the initial
adjustment phase.

The collision case trajectories, shown in panel~(b), reveal the
characteristic behavior of an overtaking interaction. The fast soliton
(amplitude $7.17 \pm 0.39$~m) propagates at a measured velocity of
0.508~m/s with $R^2 = 0.9995$, while the slow soliton (amplitude
$2.16 \pm 0.10$~m) travels at 0.185~m/s with $R^2 = 0.9950$. The
slightly reduced $R^2$ value for the slow soliton reflects the greater
relative perturbation experienced during the collision. The dashed box in
panel~(b) highlights the collision region where the trajectories
temporarily merge before separating.

Panel~(c) synthesizes the velocity--amplitude relationship across all
tracked solitons. The data points cluster along the theoretical line
$v = \varepsilon A / 3$, providing quantitative confirmation that the
computed solutions satisfy the fundamental KdV dispersion relation. The
agreement is particularly notable given that the initial conditions employ
approximate sech$^2$ profiles rather than exact soliton solutions,
demonstrating the robustness of the soliton state as an attractor in the
KdV phase space~\cite{ref-1}. This amplitude--velocity scaling is the same
relationship that governs the rank ordering and separation of internal
solitary waves observed in the ocean~\cite{ref-46,ref-48}.

Panel~(d) presents a sequence of wave profiles during the collision event,
vertically offset for visualization. The temporal progression illustrates
the collision mechanism: the faster soliton approaches from behind, the
two waves temporarily merge into a single enhanced peak, and subsequently
separate with their original amplitudes restored. This elastic scattering
behavior, wherein solitons interact nonlinearly yet emerge unchanged,
constitutes the defining characteristic that distinguishes solitons from
ordinary dispersive wave packets~\cite{ref-3}.

\begin{figure}[H]
\centering
\includegraphics[width=\textwidth]{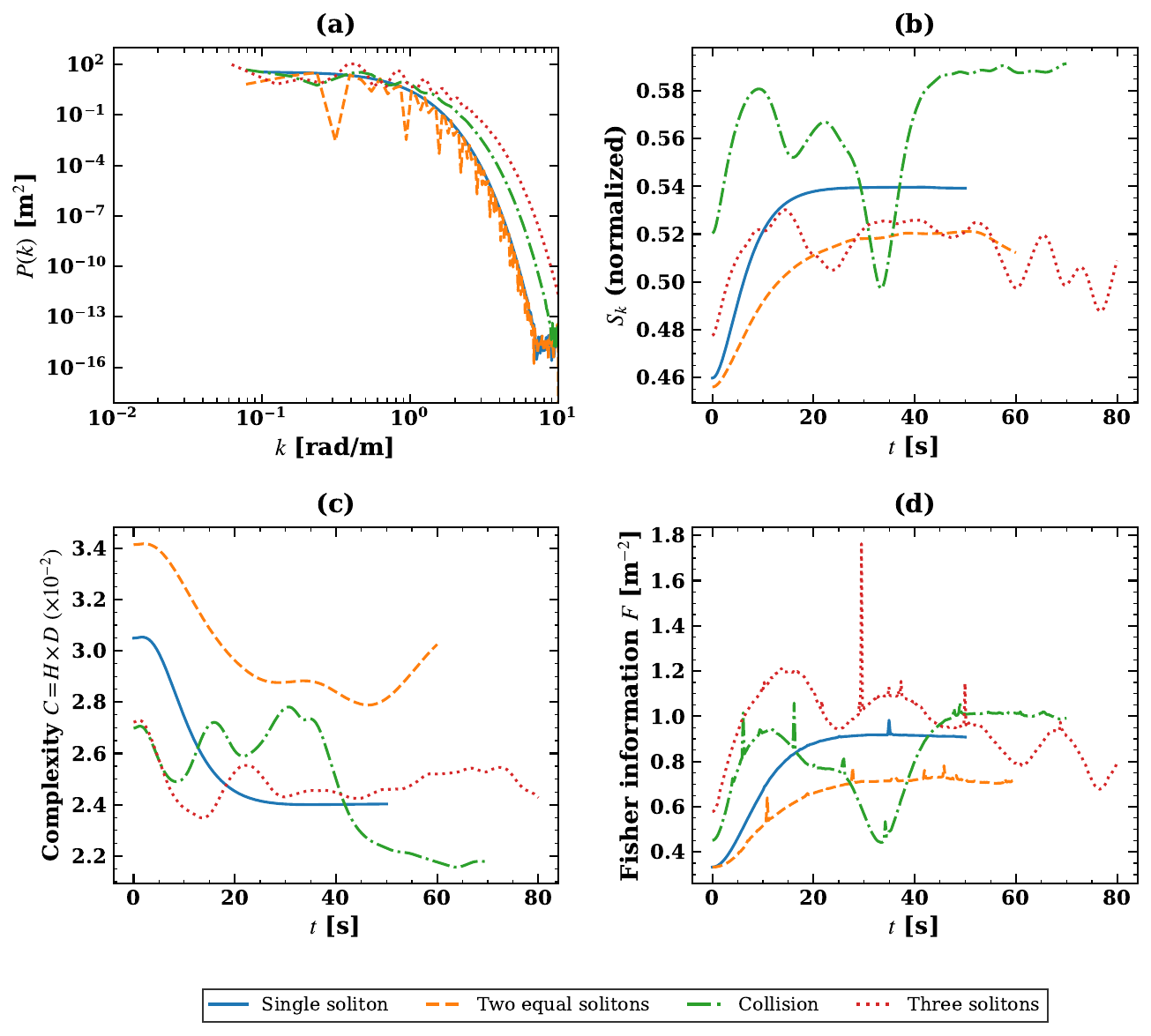}
\caption{Spectral and information-theoretic analysis.
(\textbf{a})~Power spectral density $P(k)$ at initial time for all
cases;
(\textbf{b})~normalized spectral entropy $S_k(t)$;
(\textbf{c})~LMC statistical complexity $C(t) = H \times D$;
(\textbf{d})~Fisher information $F(t)$. Line styles as in
Figure~\ref{fig:conservation}.\label{fig:spectral}}
\end{figure}

Beyond trajectory analysis, the spectral structure and information content
of the wave field provide complementary diagnostics of solution quality.
Figure~\ref{fig:spectral} presents four measures characterizing different
aspects of the spatial Fourier decomposition and its evolution.

Panel~(a) displays the initial power spectral density $P(k)$ on
logarithmic axes. All cases exhibit similar spectral structure: a plateau
at low wavenumbers ($k < 0.1$~rad/m) transitioning to exponential decay
at higher wavenumbers, consistent with the smooth sech$^2$ spatial
profiles characteristic of KdV solitons~\cite{ref-18}. The three-soliton
case (dotted red) shows the highest total power, reflecting the larger
combined amplitude, while the two equal-amplitude case (dashed orange)
exhibits oscillatory modulation arising from the spatial interference
pattern of the two separated solitons. The spectral decay extends over
approximately 15 orders of magnitude before reaching the numerical noise
floor near $10^{-16}$, confirming that aliasing errors remain negligible
for the employed spatial resolutions.

The normalized spectral entropy $S_k$, plotted in panel~(b), quantifies
the breadth of the spectral energy distribution according to
equation~\eqref{eq:spectral_entropy}. Values near unity indicate uniform
spectral content (approaching white noise), while values near zero
correspond to energy concentration at a single wavenumber (monochromatic).
The soliton solutions exhibit intermediate values in the range
$0.46$--$0.59$, reflecting their localized but smooth spatial structure.
The single-soliton case shows relatively stable entropy
($0.528 \pm 0.021$), while the collision case displays larger fluctuations
($0.567 \pm 0.024$) with transient increases during the interaction event
near $t = 35$~s. These variations suggest temporary spectral broadening
as the wave profiles distort during collision, followed by return to the
pre-collision distribution as the solitons separate---a spectral signature
of the elastic scattering process.

The LMC statistical complexity $C = H \times D$, shown in panel~(c),
provides a measure that distinguishes structured states from both trivial
order and featureless randomness~\cite{ref-37}. All cases exhibit
complexity values in the range $0.022$--$0.034$, with the two
equal-amplitude configuration showing the highest initial complexity
($C_0 = 0.034$) due to the structured interference pattern. The temporal
variations appear anticorrelated with spectral entropy: complexity
decreases as entropy increases during collision events, consistent with
the interpretation that transient spectral broadening reduces the
disequilibrium component $D$ while increasing the entropy component $H$.

Panel~(d) presents the Fisher information $F$, which measures the
sharpness or localization of the spatial amplitude distribution according
to equation~\eqref{eq:fisher_discrete}. Higher values indicate more
localized, sharper-peaked structures. The three-soliton case exhibits the
highest mean Fisher information ($0.960 \pm 0.148$~m$^{-2}$), reflecting
the presence of multiple narrow peaks, while the two equal-amplitude case
shows the lowest ($0.641 \pm 0.119$~m$^{-2}$) due to the broader,
lower-amplitude solitons. Notably, the collision case displays pronounced
spikes in Fisher information (reaching 1.8~m$^{-2}$) during the
interaction event, reflecting the transient steepening of gradients as the
soliton profiles merge. These transient features subsequently relax as the
waves separate, consistent with the elastic nature of KdV soliton
collisions.

Table~\ref{tab:spectral_summary} summarizes the mean values of the
information-theoretic measures across all test cases.

\begin{table}[H]
\centering
\caption{Summary of information-theoretic measures (mean $\pm$ standard
deviation over simulation time).\label{tab:spectral_summary}}
\small
\begin{tabular}{lccc}
\toprule
\textbf{Case} & \textbf{$S_k$} &
\textbf{$C$ ($\times 10^{-2}$)} & \textbf{$F$ [m$^{-2}$]} \\
\midrule
1 (Single soliton) & $0.528 \pm 0.021$ & $2.55 \pm 0.22$
  & $0.801 \pm 0.180$ \\
2 (Two equal)      & $0.508 \pm 0.018$ & $2.99 \pm 0.20$
  & $0.641 \pm 0.119$ \\
3 (Collision)      & $0.567 \pm 0.024$ & $2.47 \pm 0.22$
  & $0.837 \pm 0.179$ \\
4 (Three solitons) & $0.514 \pm 0.012$ & $2.49 \pm 0.07$
  & $0.960 \pm 0.148$ \\
\bottomrule
\end{tabular}
\end{table}

\begin{figure}[H]
\centering
\includegraphics[width=\textwidth]{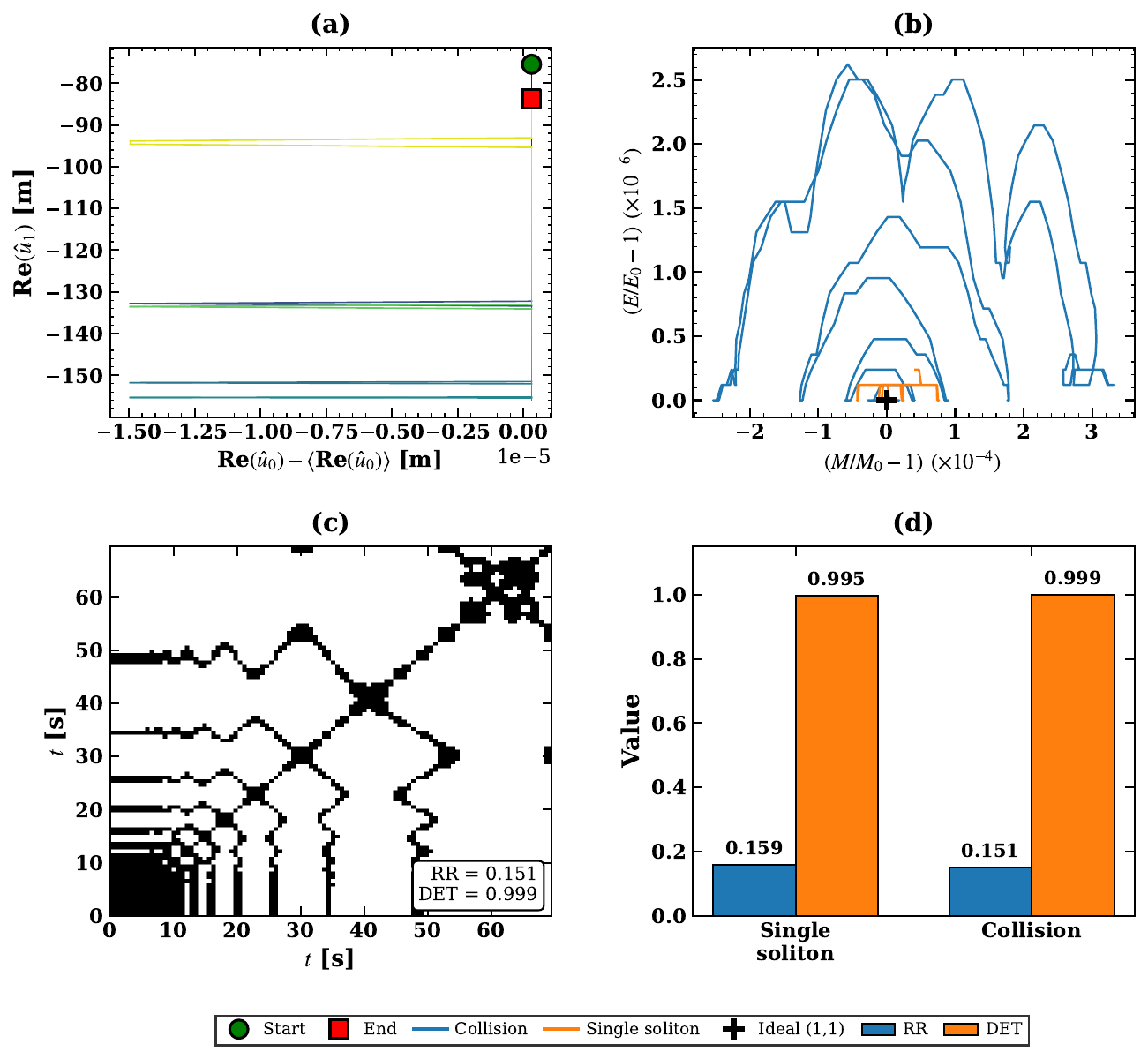}
\caption{Phase space analysis for the collision case.
(\textbf{a})~Trajectory in Fourier mode space (real parts of modes 0 and
1), colored by time;
(\textbf{b})~trajectory in normalized conservation space
$(M/M_0, E/E_0)$;
(\textbf{c})~recurrence plot with quantification metrics;
(\textbf{d})~comparison of recurrence rate (RR) and determinism (DET)
between single-soliton and collision
cases.\label{fig:phase_space}}
\end{figure}

The complete integrability of the KdV equation implies that phase space
trajectories are confined to low-dimensional invariant
manifolds~\cite{ref-5}. RQA provides tools to characterize this structure
without requiring explicit identification of action-angle variables.
Figure~\ref{fig:phase_space} presents phase space diagnostics for the
collision case (Case~3) alongside comparison with the single-soliton
reference.

Panel~(a) displays the trajectory in a two-dimensional projection of
Fourier mode space, plotting the real part of the first mode
$\text{Re}(\hat{u}_1)$ against deviations of the zeroth mode
$\text{Re}(\hat{u}_0) - \langle \text{Re}(\hat{u}_0) \rangle$. The
zeroth mode, proportional to the spatial mean, exhibited negligible
variation ($|\hat{u}_0| = 178.85 \pm 0.00$~m), consistent with mass
conservation. Higher modes showed larger fluctuations correlated with
collision dynamics: $|\hat{u}_1| = 149.9 \pm 16.0$~m and
$|\hat{u}_2| = 97.8 \pm 27.2$~m. The trajectory traces a bounded path
in this projection, with the color gradient indicating temporal
progression from the start (green circle) to end (red square) of the
simulation.

Panel~(b) presents the trajectory in normalized conservation space,
plotting $(M/M_0 - 1)$ against $(E/E_0 - 1)$. The ideal conservative
dynamics would confine the trajectory to a single point at the origin.
The computed trajectory remains confined to an extremely small region:
$M/M_0 \in [0.9997, 1.0003]$ and $E/E_0 \in [1.0000, 1.0000026]$, with
total extent approximately $5.86 \times 10^{-4}$. The single-soliton
case (orange trace) occupies an even smaller region, while both
trajectories cluster near the ideal point (black cross), confirming the
high fidelity of the numerical conservation.

The recurrence plot in panel~(c) visualizes the binary recurrence matrix
$R_{ij}$ defined in equation~\eqref{eq:recurrence_matrix}, where black
pixels indicate times $t_i$ and $t_j$ at which the conservation space
trajectory revisits the same neighborhood (within threshold
$\varepsilon = 3.52 \times 10^{-5}$). The diagonal line structures
indicate deterministic dynamics: when the system visits a particular
state, it subsequently follows a trajectory parallel to previous visits,
characteristic of regular (non-chaotic) motion on invariant
manifolds~\cite{ref-41,ref-42}. The recurrence rate RR $= 0.151$
indicates that approximately 15\% of state pairs satisfy the recurrence
criterion, while the determinism DET $= 0.999$ confirms that nearly all
recurrences form diagonal line structures.

Panel~(d) compares the RQA metrics between the single-soliton and
collision cases. Both exhibit similar recurrence rates (0.159 and 0.151,
respectively) and high determinism (0.995 and 0.999). The marginally
higher determinism in the collision case may appear counterintuitive but
likely reflects the longer simulation time (70~s versus 50~s) providing
more opportunities for trajectory segments to align. A correlation
dimension estimate of 0.86 for the collision case, computed from the
recurrence matrix scaling~\cite{ref-42}, suggests that the effective
dynamics occupy a manifold of dimension less than unity---consistent with
the trajectory being confined near a fixed point (the ideal conservation
state) with small bounded oscillations.

\section{Discussion}

The results presented above demonstrate that the \texttt{sangkuriang}
solver successfully reproduces the essential features of KdV soliton
dynamics while maintaining numerical accuracy appropriate for research
applications in idealized ocean wave modeling. Several aspects merit
further discussion in the context of GFD and coastal oceanography.

The pseudo-spectral spatial discretization combined with eighth-order
adaptive time integration achieved conservation errors ranging from
$\mathcal{O}(10^{-7})$ to $\mathcal{O}(10^{-4})$ across the test cases.
These values compare favorably with those reported in the literature for
similar methods applied to nonlinear dispersive equations arising in ocean
wave theory~\cite{ref-17,ref-20}. The observation that momentum and energy
are conserved more accurately than mass warrants consideration. The mass
functional~\eqref{eq:mass_invariant} involves only the solution field
itself, making it sensitive to any drift in the solution mean. In
contrast, the momentum functional~\eqref{eq:momentum_invariant} depends on
$u^2$, which suppresses the contribution of small-amplitude dispersive
radiation relative to the dominant soliton peaks. The energy
functional~\eqref{eq:energy_invariant}, involving spatial derivatives,
further weights the smooth soliton cores over high-wavenumber numerical
artifacts. This hierarchy suggests that the mass error may serve as a
conservative upper bound on overall numerical accuracy. From an
oceanographic standpoint, the high fidelity of momentum and energy
conservation is particularly reassuring, as these quantities govern the
transport of wave energy and the forces exerted by solitary waves on
offshore structures and the seabed~\cite{ref-48}.

The absence of systematic drift in the conservation time series indicates
that the DOP853 integrator, while not symplectic, introduces primarily
oscillatory rather than secular errors for the integration times
considered. For significantly longer simulations---such as those required
to model solitary wave propagation across an entire continental shelf or
ocean basin---the accumulation of these oscillatory errors could
potentially become problematic, and structure-preserving
integrators~\cite{ref-33,ref-43} might offer advantages despite their
typically higher computational cost per step.

The close agreement between measured and theoretical soliton velocities
($< 5\%$ discrepancy) confirms that the computed solutions satisfy the
fundamental amplitude--velocity relationship~\eqref{eq:soliton_velocity}.
This nonlinear dispersion relation is the same scaling that governs the
rank ordering and separation of internal solitary waves observed in the
ocean, where larger-amplitude waves consistently propagate faster than
their smaller counterparts~\cite{ref-46,ref-49}. The small systematic
excess in measured velocities relative to predictions based on mean
amplitude may arise from several factors. First, the soliton amplitudes
evolve during the initial adjustment phase as the approximate initial
conditions relax toward exact soliton profiles, and velocity measurements
based on trajectory slopes integrate over this transient. Second, the
theoretical velocity assumes an isolated soliton in an infinite domain,
whereas the periodic boundary conditions and finite domain introduce
corrections that may slightly modify the effective dispersion
relation~\cite{ref-18}. Similar discrepancies between observed and
theoretically predicted propagation speeds have been documented in field
studies of oceanic internal solitary waves, where background currents,
stratification variability, and bathymetric effects introduce additional
corrections to the idealized KdV prediction~\cite{ref-47,ref-50}.

The high linearity of all tracked trajectories ($R^2 > 0.99$) provides
strong evidence that the computed waves behave as true solitons rather
than dispersive wave packets that would spread and decelerate over time.
This validation is particularly meaningful for the post-collision
segments, where any deviation from integrability would manifest as
amplitude-dependent velocity changes. In the oceanic context, the
persistence of soliton identity through collisions is a key feature that
enables the long-range propagation of internal solitary waves across
hundreds of kilometers of continental shelf~\cite{ref-46,ref-51}.

The spectral and information-theoretic analyses reveal consistent patterns
that reflect the integrable structure of the KdV equation. The bounded
variations of spectral entropy, statistical complexity, and Fisher
information---with transient excursions during collisions that subsequently
relax---indicate that the wave field returns to its pre-collision
information content. This behavior contrasts with what would be expected
for non-integrable systems, where collision events could transfer energy to
radiation modes, leading to secular changes in these
measures~\cite{ref-6}. In realistic ocean settings, perturbations to
perfect integrability arise from variable stratification, background shear,
and bathymetric gradients~\cite{ref-50}; the diagnostic tools developed
here could serve to quantify the degree to which such perturbations disrupt
the regular structure of the idealized KdV dynamics.

The Fisher information proved particularly sensitive to collision dynamics,
with pronounced spikes corresponding to the transient steepening of
gradients during wave merging. This sensitivity suggests that Fisher
information may serve as a useful diagnostic for detecting the onset and
completion of soliton interactions in more complex scenarios, including
the multi-wave interaction events observed in energetic oceanic internal
wave environments such as the South China Sea~\cite{ref-49} and the Strait
of Gibraltar.

The RQA provides independent confirmation of the regular, non-chaotic
dynamics expected for an integrable Hamiltonian system. The high
determinism values (DET $> 0.99$) indicate that the phase space trajectory
exhibits the parallel structure characteristic of motion on invariant tori,
rather than the mixing behavior associated with chaotic
dynamics~\cite{ref-42}. The estimated correlation dimension below unity is
consistent with the trajectory being confined near a fixed point (the
exact conservation state) with small-amplitude oscillations.

These results suggest that RQA could serve as a model-independent
diagnostic for assessing the integrability of computed solutions in
situations where analytical results are unavailable. Significant
deviations from high determinism might indicate either numerical breakdown
of integrability due to insufficient resolution, or genuine non-integrable
dynamics in perturbed KdV-type equations. This capability is potentially
valuable for ocean wave modeling, where extended KdV equations with
variable coefficients are routinely employed to account for realistic
oceanographic conditions~\cite{ref-48,ref-50}, and where the degree to
which the integrable structure survives such extensions is not always
known a priori.

The demonstration that research-grade KdV simulations can be performed in
minutes on a laptop-class workstation represents a practical advantage for
exploratory studies in physical oceanography and GFD. The
Numba-accelerated implementation achieved throughputs of approximately
1000--1400 time steps per second, suggesting that the solver is not
severely memory-bandwidth limited for the grid sizes considered. The
observed scaling from $N = 512$ to $N = 1024$ (throughput reduction from
$\sim$1300 to $\sim$1030 steps/s) is consistent with the
$\mathcal{O}(N \log N)$ FFT complexity combined with increased cache
pressure. The accessibility of this computational tool on commodity
hardware lowers the barrier for researchers and graduate students in
oceanography to conduct idealized numerical experiments on nonlinear
dispersive wave dynamics without requiring institutional high-performance
computing allocations.

For applications requiring higher resolution or longer integration
times---such as simulating solitary wave propagation over shelf-scale
distances or conducting ensemble studies across a range of oceanic
stratification profiles---several optimization strategies could be
considered: vectorized evaluation of multiple initial conditions in
parallel, GPU acceleration of FFT operations via CuPy~\cite{ref-44}, or
time-parallel algorithms for ensemble simulations. The modular
architecture of \texttt{sangkuriang} is designed to facilitate such
extensions while maintaining compatibility with the existing analysis
framework.

Several limitations of the present study should be acknowledged. First,
the test cases employed approximate sech$^2$ initial conditions rather
than exact $N$-soliton solutions obtainable from inverse scattering
theory~\cite{ref-5}; comparison with exact solutions would provide a more
stringent validation. Second, the periodic boundary conditions preclude
investigation of soliton interactions with boundaries or non-periodic wave
backgrounds, which are relevant to realistic coastal settings where waves
encounter shorelines, headlands, or river plumes. Third, the standard KdV
equation considered here represents only the leading-order balance between
nonlinearity and dispersion; extensions to higher-order equations (e.g.,
the modified KdV or Kawahara equations~\cite{ref-45}) and
variable-coefficient formulations that account for slowly varying
bathymetry and stratification~\cite{ref-50} would broaden the solver's
applicability to different oceanic regimes. Fourth, the present study
considers only surface wave analogs; the extension to internal wave
dynamics in continuously stratified fluids requires the incorporation of
modal structure and density-dependent coefficients~\cite{ref-46,ref-48}.

Future development of \texttt{sangkuriang} could address these limitations
while maintaining the design philosophy of accessibility and
reproducibility. The integration of inverse scattering routines would
enable exact initial conditions for validation studies. Variable-coefficient
extensions would permit modeling of wave propagation over varying
bathymetry, a critical capability for realistic coastal ocean
applications~\cite{ref-50,ref-51}. The incorporation of rotational effects
through the Ostrovsky equation would extend applicability to internal waves
at scales where the Earth's rotation becomes significant~\cite{ref-47}.
Perhaps most significantly, the pseudo-spectral framework could be extended
to two-dimensional generalizations such as the Kadomtsev--Petviashvili
equation~\cite{ref-6}, enabling investigation of oblique soliton
interactions and transverse instabilities relevant to the three-dimensional
propagation patterns of oceanic internal waves observed in satellite
imagery~\cite{ref-49}.

\section{Conclusions}

We have presented \texttt{sangkuriang}, an open-source pseudo-spectral
solver for the KdV equation combining Fourier-based spatial discretization
with adaptive eighth-order Runge--Kutta time integration. Systematic
validation across four test cases of increasing complexity---from isolated
soliton propagation to three-body interactions motivated by oceanic
internal solitary wave dynamics---confirmed conservation of the KdV
invariants to relative errors below $10^{-5}$ for momentum and energy,
while measured soliton velocities agreed with the theoretical
amplitude--velocity relation to within 5\%. Information-theoretic
diagnostics revealed bounded spectral entropy and near-unit phase-space
determinism throughout all simulations, providing quantitative evidence
that the numerical solutions preserve the integrable Hamiltonian structure
of the governing equation. Implemented in Python with Numba-accelerated
parallel execution, the solver completed the most demanding
configuration---a three-soliton interaction over 80~s at $N = 1024$
resolution---in under ten minutes on a laptop workstation, demonstrating
accessibility for researchers in oceanography and GFD without requiring
high-performance computing resources.

The idealized test cases examined here---single-soliton propagation,
symmetric pairs, overtaking collisions, and rank-ordered three-wave
packets---capture the essential nonlinear dispersive dynamics that underpin
a wide range of oceanic solitary wave phenomena, from internal soliton
trains on continental shelves to long gravity waves in coastal basins. By
establishing the fidelity of the \texttt{sangkuriang} solver against these
canonical benchmarks, the present work provides a verified computational
foundation upon which more realistic ocean wave studies can be built.

Beyond its immediate utility for KdV simulations, the diagnostic framework
developed here---combining conservation monitoring, spectral information
measures, and recurrence quantification---offers a transferable methodology
for assessing integrability preservation in numerical solutions of nonlinear
dispersive PDEs arising in GFD. Extensions to variable-coefficient KdV
variants relevant to wave propagation over realistic bathymetry,
higher-order dispersive equations such as the Kawahara equation, rotational
generalizations including the Ostrovsky equation for internal waves
influenced by the Earth's rotation, and two-dimensional formulations such
as the Kadomtsev--Petviashvili equation represent natural directions for
future development within the existing pseudo-spectral framework. Such
extensions would progressively bridge the gap between the idealized
dynamics studied here and the full complexity of nonlinear wave propagation
in the real ocean.

\section*{Author Contributions}

Conceptualization, D.E.I. and S.H.S.H.; methodology, D.E.I., S.H.S.H.,
and F.K.; software, S.H.S.H.; validation, D.E.I., S.H.S.H., and F.K.;
formal analysis, S.H.S.H.; investigation, D.E.I. and S.H.S.H.; resources,
D.E.I., I.P.A., and R.S.; data curation, S.H.S.H.; writing---original
draft preparation, S.H.S.H.; writing---review and editing, D.E.I., A.P.,
R.D.K., F.K., I.P.A., K.A.S., A.P.H., F.R.F., and R.S.; visualization,
S.H.S.H.; supervision, D.E.I., I.P.A., K.A.S., A.P.H., F.R.F., and R.S.;
project administration, D.E.I.; funding acquisition, D.E.I., S.H.S.H.,
F.K., A.P.H., F.R.F., and R.S. All authors have read and agreed to the
published version of the manuscript.

\section*{Funding}

This study was supported by the Program Penelitian, Pengabdian kepada
Masyarakat, dan Inovasi FITB 2026, the Dean's Distinguished Fellowship
2023 from the University of California, Riverside, the Riset Dosen Muda
2025, the Riset Unggulan 2026, and the Riset Talenta Unggul 3P 2026 from
Bandung Institute of Technology.

\section*{Data Availability}

The \texttt{sangkuriang} library source code is available on GitHub at\\
\url{https://github.com/sandyherho/sangkuriang-ideal-solver} under the MIT
license and can be installed from PyPI at
\url{https://pypi.org/project/sangkuriang-ideal-solver/}. The Python
scripts for data analysis
(\url{https://github.com/sandyherho/suppl_sangkuriang}) and all simulation
outputs---including animations, NetCDF data files, figures, computational
logs, and statistical analyses---archived on the Open Science Framework
(OSF) (\url{https://doi.org/10.17605/OSF.IO/MS5EJ}) are released under
the MIT license.

\section*{Acknowledgments}

The authors thank the anonymous reviewers for their constructive comments.
During the preparation of this work, the authors used Claude Sonnet 4.5
(Anthropic) to assist with English grammar, vocabulary, and stylistic
refinement. All mathematical derivations, numerical methods, scientific
analysis, interpretation of results, and intellectual content were produced
entirely by the authors. The authors reviewed and edited all AI-assisted
outputs and take full responsibility for the content of the publication.

\section*{Conflicts of Interest}

The authors declare no conflicts of interest. The funders had no role in
the design of the study; in the collection, analyses, or interpretation of
data; in the writing of the manuscript; or in the decision to publish the
results.

\section*{Abbreviations}

\noindent
\begin{tabular}{@{}ll}
CF   & Climate and Forecast\\
CPU  & Central Processing Unit\\
DFT  & Discrete Fourier Transform\\
DOP853 & Dormand--Prince 8(5,3) method\\
FFT  & Fast Fourier Transform\\
FFTW & Fastest Fourier Transform in the West\\
GFD  & Geophysical Fluid Dynamics\\
IFFT & Inverse Fast Fourier Transform\\
ISW  & Internal Solitary Wave\\
JIT  & Just-In-Time\\
KdV  & Korteweg--de Vries\\
LLVM & Low Level Virtual Machine\\
LMC  & L\'{o}pez--Ruiz--Mancini--Calbet\\
MKL  & Math Kernel Library\\
NetCDF & Network Common Data Format\\
ODE  & Ordinary Differential Equation\\
OpenMP & Open Multi-Processing\\
PDE  & Partial Differential Equation\\
PyPI & Python Package Index\\
RMS  & Root Mean Square\\
RQA  & Recurrence Quantification Analysis
\end{tabular}

\end{document}